\documentclass[12pt]{article}

\usepackage[export]{adjustbox}

\usepackage{cite,graphicx,amsmath,amssymb}
\usepackage{multicol}
\usepackage{xcolor}
\usepackage{slashed}
\usepackage[normalem]{ulem}
\usepackage{bbm}
\usepackage{setspace}
\usepackage{footmisc}
\usepackage{hyperref}
\usepackage{cancel}
\usepackage{mathtools}

\usepackage{bbm}
\usepackage{MnSymbol}

\usepackage{booktabs}
\usepackage{multirow}
\usepackage{arydshln}

\usepackage{subfig}

\graphicspath{{./Pictures/}}

\makeatletter
\renewcommand{\fnum@figure}{Fig. \thefigure}
\renewcommand{\fnum@table}{Tab. \thetable}
\makeatother

\definecolor{links}{rgb}{0.1,0.3,1}
\definecolor{cites}{rgb}{0.2,0.5,0.1}
\definecolor{urls}{RGB}{0,139,139}
\hypersetup{
	bookmarks=true,         
	unicode=false,          
	pdftoolbar=true,        
	pdfmenubar=true,        
	pdffitwindow=true,     
	pdfstartview={FitH},    
	pdfsubject={},   
	pdfcreator={},   
	pdfproducer={}, 
	pdfnewwindow=true,      
	colorlinks=true,       
	linkcolor=links,          
	citecolor=cites,        
	filecolor=magenta,      
	urlcolor=urls,           
	breaklinks=true,		
}

\newcommand{\be}{\begin{equation}}
\newcommand{\ee}{\end{equation}}

\newcommand{\vev}[1]{\left\langle#1\right\rangle}
\renewcommand{\d}{\text{d}}

\newcommand{\fieldlable}[1]{\text{\tiny#1}}
\newcommand{\actionlabel}[1]{\text{\tiny #1}}
\newcommand{\dx}[2]{\text{d}^{#1}\hspace{-1pt}#2}
\newcommand{\M}{\mathcal{M}}
\newcommand{\W}{\mathcal{W}}
\newcommand{\V}{\mathcal{V}}

\newcommand{\Mpl}{M_{\text{Pl}}}                

\newcommand{\gs}{g_\text{s}}				    
\newcommand{\ls}{l_\text{s}}                    
\newcommand{\stsc}{M_\text{s}} 		            
\newcommand{\vol}[1]{\text{vol}(#1)} 			
\newcommand{\intmfd}{\mathcal{X}^{6}} 				
\newcommand{\extmfd}{\mathcal{M}^{1,3}} 		
\newcommand{\kmp}{\chi}                         
\newcommand{\bt}{T}			                        
\newcommand{\ap}{\alpha'}                           
\newcommand{\hs}[1]{\star_{\tiny #1}}               

\newcommand{\whs}[1]{\wedge\hspace{-2pt}\hs{#1}}    
 
\newcommand{\tr}{\text{tr}} 
\newcommand{\codiff}{\text{d}^\dagger}

\renewcommand{\Im}{\text{Im}}
\renewcommand{\Re}{\text{Re}}
\newcommand{\bcgr}[1]{\bar{#1}}                     

\newcommand{\dfour}{\d^{\fieldlable{(4)}}}
\newcommand{\ddfour}{\d^{\fieldlable{(4)}\dagger}}
\newcommand{\dsix}{\d^{\fieldlable{(6)}}}
\newcommand{\ddsix}{\d^{\fieldlable{(6)}\dagger}}
\newcommand{\pair}[1]{\underline{#1}}

\newcommand{\ampl}[1]{{\cal A}(\text{\ref{#1}})}

\begin{document}

\thispagestyle{empty}

\begin{center}

{
\Large  
Small Kinetic Mixing in String Theory
}
\\[2cm]

{\large  Arthur Hebecker$^{\,a,}$\footnote{\href{mailto:a.hebecker@thphys.uni-heidelberg.de}{a.hebecker@thphys.uni-heidelberg.de}}, Joerg Jaeckel$^{\,a,}$\footnote{\href{mailto:jjaeckel@thphys.uni-heidelberg.de}{jjaeckel@thphys.uni-heidelberg.de}} and Ruben Kuespert$^{\,a,b,}$\footnote{\href{mailto:r.kuespert@thphys.uni-heidelberg.de}{r.kuespert@thphys.uni-heidelberg.de}}}

\vspace{1cm}

{\it

${}^{a}$ Institut f\"ur theoretische Physik, Universit\"at Heidelberg, Philosophenweg 16 / 19,\\69120 Heidelberg, Germany\\[.1cm]
${}^{b}$ Max-Planck-Institut f\"ur Kernphysik, Saupfercheckweg 1,\\69117 Heidelberg, Germany\\[1.6cm]
}


{\bf Abstract}
\end{center}
Kinetic mixing between gauge fields of different $U(1)$ factors is a well-studied phenomenon in 4d EFT. In string compactifications with $U(1)$s from sequestered D-brane sectors, kinetic mixing becomes a key target for the UV prediction of a phenomenologically important EFT operator. Surprisingly, in many cases kinetic mixing is absent due to a non-trivial cancellation. In particular, D3-D3 kinetic mixing in type-IIB vanishes while D3-anti-D3 mixing does not. This follows both from exact CFT calculations on tori as well as from a leading-order 10d supergravity analysis, where the key cancellation is between the $C_2$ and $B_2$ contribution. We take the latter approach, which is the only one available in realistic Calabi-Yau settings, to a higher level of precision by including sub-leading terms of the brane action and allowing for non-vanishing $C_0$. The exact cancellation persists, which we argue to be the result of $SL(2,\mathbb{R})$ self-duality.
We note that a $B_2C_2$ term on the D3-brane, which is often missing in the recent literature, is essential to obtain the correct zero result. Finally, allowing for $SL(2,\mathbb{R})$-breaking fluxes, kinetic mixing between D3-branes arises at a volume-suppressed level. We provide basic explicit formulae, both for kinetic as well as magnetic mixing, leaving the study of phenomenologically relevant, more complex situations for the future. We also note that describing our result in 4d supergravity appears to require higher-derivative terms -- an issue which deserves further study.

\newpage

\hypersetup{linkcolor=black}
\addtocontents{toc}{\protect\hypertarget{toc}{}}
\tableofcontents 
\hypersetup{linkcolor=links}
\newpage


\section{Introduction}\label{sc:intro}
\setcounter{footnote}{0}
Kinetic mixing (KM)~\cite{Okun:1982xi, Holdom:1985ag} is one of the four portals \cite{Antel:2023hkf} coupling the Standard Model (SM) to a possible hidden sector which, in particular, may contain
dark matter (DM). For KM to be relevant, 
the hidden sector should include a U(1) gauge group, usually referred to as a dark or hidden U(1). 

Let us denote the field strength tensor of the visible or Standard-Model U(1) by $\mathbb{F}_\fieldlable{(A)}=\d \mathbb{A}_\fieldlable{(A)}$ and that of the hidden U(1) by $\mathbb{F}_\fieldlable{(B)}=\d \mathbb{A}_\fieldlable{(B)}$. We take these fields to point parallel to the axes of the dual charge lattice and to be normalized such that their $\mathbb{F}_\fieldlable{(A/B)}^2$ terms have the canonical prefactor $-1/4$. Generically\footnote{
An exception arises if a charge conjugation symmetry $\cal C$, acting only on one sector and leaving the other sector invariant, is present \cite{Dienes:1996zr, Garny:2018grs, Gherghetta:2019coi}. Specifically, the transformation properties 
${\cal C}[\mathbb{A}^\fieldlable{(A)}]=\mathbb{A}^\fieldlable{(A)}$ and ${\cal C}[\mathbb{A}^\fieldlable{(B)}]=-\mathbb{A}^\fieldlable{(B)}$ obviously exclude a KM term.
},
kinetic~\cite{Okun:1982xi,Holdom:1985ag} and magnetic~\cite{Brummer:2009cs, Brummer:2009oul} mixing operators will then also be present in the Lagrangian,
\begin{equation}
    \label{eq:kinetic-mixing-def}
\mathcal{L} \supset -\frac{\kmp_\fieldlable{AB}}{2} \mathbb{F}_\fieldlable{(A)}^{\mu\nu} \mathbb{F}^\fieldlable{(B)}_{\mu\nu} -\frac{\tilde{\kmp}_\fieldlable{AB}}{2} \mathbb{F}_\fieldlable{(A)}^{\mu\nu} \tilde{\mathbb{F}}^\fieldlable{(B)}_{\mu\nu} ~,
\end{equation}
with $\kmp_\fieldlable{AB}$ and $\tilde{\kmp}_\fieldlable{AB}$ the kinetic and magnetic mixing parameters.

\subsection{Small kinetic mixing in type IIB string compactifications}
\begin{figure}[t]
    \centering
    \subfloat[]{\includegraphics[width=0.44\textwidth,valign=c]{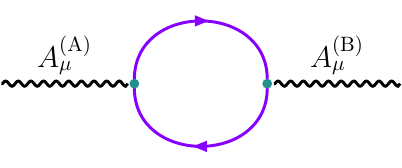}
    \vphantom{\includegraphics[width=0.44\textwidth,valign=c]{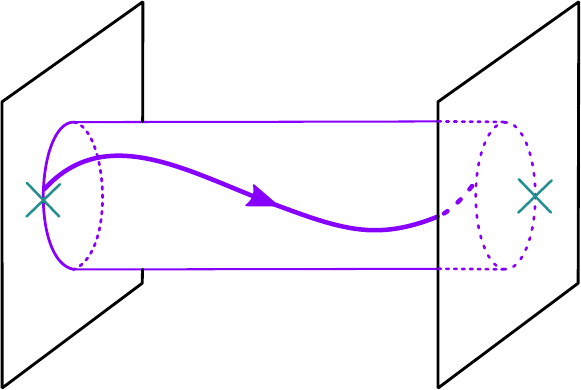}}
    \label{fig:KM-diagrams-a}}
    \hspace{30pt}
    \subfloat[]{\includegraphics[width=0.44\textwidth,valign=c]{Pictures/StringLoop.pdf}
    \label{fig:KM-diagrams-b}}
    \caption{Origin of KM in field and string theory. (a) A particle charged under both $U(1)_{\fieldlable{(A)}}$ and $U(1)_{\fieldlable{(B)}}$ runs in the loop. (b) Stringy analog of (a) where the charged particle is replaced by a string stretched between D-branes supporting the two $U(1)$ gauge theories.}
    \label{fig:KM-diagrams}
\end{figure}

As we will discuss in sect.~\ref{pmot}, KM has to be extremely small to avoid experimental detection and be of phenomenological interest. Typical values are as low as $\kmp_\fieldlable{AB}\sim 10^{-15}$. This has to be contrasted with the field-theoretic expectation that KM is a 1-loop effect, 
cf.~fig.~\ref{fig:KM-diagrams-a}. By the Completeness Conjecture \cite{Polchinski:2003bq,Banks:2010zn} or, more quantitatively \cite{Benakli:2020vng, Obied:2021zjc} by the Weak Gravity Conjecture~\cite{Arkani-Hamed:2006emk}, 
heavy states $\Phi$ charged under $U(1)_\fieldlable{(A)}$ and $U(1)_\fieldlable{(B)}$ should always be present. This leads to the estimate
\begin{equation}
     \label{eq:KMparameter}
 	\kmp_\fieldlable{AB} \sim c_{\text{loop}}~ g_\fieldlable{(A)} g_\fieldlable{(B)}~\ln\left( \frac{\Lambda^2}{m^2_\Phi} \right)\,,
\end{equation}
where $\Lambda$ is the EFT cutoff and $c_{\text{loop}}\sim 1/16\pi^2$ the loop suppression factor. Clearly, the suppression by
$c_{\text{loop}}$ is insufficient for phenomenological purposes. One needs either a tiny hidden gauge coupling $g_\fieldlable{(B)}$ or an overwhelmingly precise cancellation between the loop effects of different charged states $\Phi$.

String theory has the potential to produce such a precise cancellation. This may be viewed as providing `loopholes' in the generic field theory prediction stated above~\cite{Obied:2021zjc}. We will argue that some of the most natural and well-studied settings allow for such loopholes and predict surprisingly small mixing parameters.
In other words, our results suggest that the above estimate \eqref{eq:KMparameter} can be very misleading in stringy scenarios. Instead, when we summarize our findings in sect.~\ref{sc:pheno} we obtain a volume suppression of $\kmp_\fieldlable{AB}$,
\begin{equation}
    \kmp_\fieldlable{AB} \sim \V^{-4/3} ~,\label{v43}
\end{equation}
resulting in tiny values, see fig.~\ref{fig:KM-Parameter-Plots} below.
We emphasise that these tiny values of $\kmp_\fieldlable{AB}$ arise without small gauge couplings and without tuning -- they are simply the result of the sequestering of the respective sectors in the compact dimensions.
That said, and as we will discuss in more detail in sect.~\ref{sc:issues-with-sugra}, this result is puzzling from a 4d supergravity point of view:
A priori, holomorphicity of the gauge-kinetic function combined with the shift symmetry of Kahler moduli prevents a volume dependence of the form \eqref{v43}.\footnote{We would like to thank Joe Conlon for reminding us of this issue. See also~\cite{Goodsell:2009xc,Bullimore:2010aj}.} We discuss in sect.~\ref{sc:issues-with-sugra} possibilities of how this can be resolved.

\begin{figure}[h]
    \centering
    \subfloat[]{\includegraphics[height=2.5cm,valign=c]{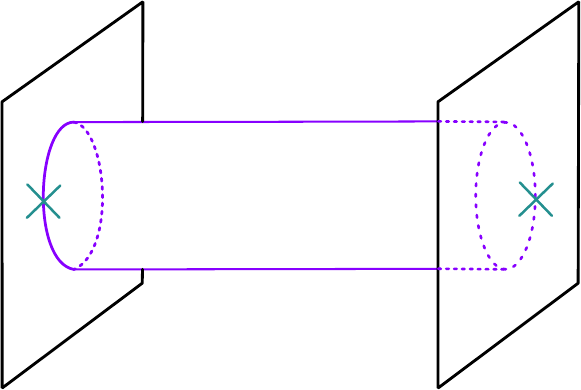}
    \label{fig:KM-explanation-a}
    }
    \hspace{13pt}
    \subfloat[]{\includegraphics[height=2.5cm,valign=c]{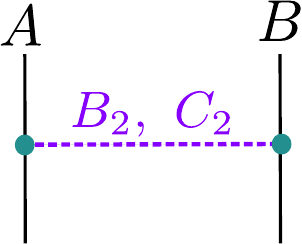}
    \label{fig:KM-explanation-b}
    }
    \hspace{13pt}
    \subfloat[]{\includegraphics[width=0.35\textwidth,valign=c]{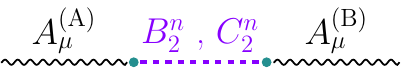}
    \vphantom{\includegraphics[height=2.5cm,valign=c]{Pictures/StringLoop.pdf}}
    \label{fig:KM-explanation-c}
    }
    \caption{Different perspectives on KM: (a) shows the string loop diagram; (b) is its reinterpretation as exchange of closed string fields in 10d supergravity; (c) symbolizes the corresponding 4d point-of-view, where the closed-string effect is encoded in a tower of KK modes (labeled by $n$) which mix with the gauge fields on the two branes.}
    \label{fig:KM-explanation}
\end{figure}

To see this, let us first recall the relation between field-theory loop and string loop calculations \cite{Dienes:1996zr,Abel:2003ue}: We focus on type IIB, where gauge theories live on D-branes and the relevant string loop diagram 
is the cylinder shown in fig.~\ref{fig:KM-diagrams-b}. Field-theoretically, one may think of this cylinder as of a heavy string state, stretched between the branes, and running in the loop. At large brane separation, it is advantageous to appeal to open-closed string duality and reinterpret the cylinder as the tree-level exchange of massless 10d fields between the branes \cite{Abel:2008ai, Goodsell:2009pi, Goodsell:2010ie, Cicoli:2011yh},
see fig.~\ref{fig:KM-explanation}. Our focus in the rest of the paper is on D3-branes. In this case, the relevant fields are the Kalb-Ramond field $B_2$ and the Ramond-Ramond field $C_2$.\footnote{
For 
different branes other $C_p$ forms also contribute.
}
Crucially, in the phenomenologically most interesting case of O3/O7 orientifold models, $B_2/C_2$ have no Kaluza-Klein (KK) zero modes such that integrating out the KK tower gives local 4d mixing terms as in eq.~\eqref{eq:kinetic-mixing-def}.  Further references to ideas in the context of KM with extra dimensions, also outside of string theory, can be found in~\cite{Burgess:2008ri,Benakli:2009mk,Gherghetta:2010cq,McDonald:2010iq,Heckman:2010fh,Bullimore:2010aj,McDonald:2010fe,Goodsell:2011wn,Camara:2011jg,Marchesano:2014bia,Jaeckel:2014eba,Rizzo:2018ntg,DeOliveiraJunior:2019mrh,Rizzo:2020ybl,Anastasopoulos:2020xgu,Wojcik:2022rtk}.

Of course, one may try to choose a regime of small gauge couplings to suppress mixing. However, this is essentially impossible in the D3 case. Even in the D7 case, where large brane volumes can in principle be considered, one runs into phenomenological limits \cite{Goodsell:2009xc, Cicoli:2011yh, Benakli:2020vng}. More interesting is the fact that, if the relevant branes are separated within a large-volume Calabi-Yau, then $B_2/C_p$ propagation is suppressed and mixing becomes small due to sequestering. This is obvious when thinking in terms of tree-level propagation, but it is very surprising from the 4d loop perspective. In this latter way of thinking, a precise cancellation between the contributions from many open string states occurs. However, as has also been known for a long time, the suppression can become even much stronger: In the D$p$-D$p$-brane case, an exact cancellation leads to vanishing mixing at leading order. Specifically for the mixing between two $D3$ branes, which is our main focus, this exact cancellation occurs between the contributions of $B_2$ and $C_2$ \cite{Abel:2003ue, Abel:2008ai, Goodsell:2009xc}.

Our goal in the present paper is a better understanding of this mysterious but phenomenologically very important cancellation. We go beyond a leading-order analysis, aiming at either the proof of an all-orders zero result or at the identification of the leading, non-zero mixing effect.
We start in sect.~\ref{sc:SL2R-notations} by introducing our notation and  presenting the key formulae for an SL$(2,\mathbb{R})$-covariant treatment of the D3 action. In sect.~\ref{sc:LOC}, we re-derive the leading-order cancellation from \cite{Abel:2008ai} in a way suitable for the following generalizations.
Next, in sect.~\ref{sc:GCwoF}, we extend this analysis to include two key sub-leading effects: a non-zero background value of $C_0$ as well as the self-couplings and mixing of $B_2$ and $C_2$ in the D3-brane action. On the basis of a calculation which crucially relies on the SL$(2,\mathbb{R})$ structure, we find that the exact cancellation persists. 
An essential prerequisite for this non-trivial result 
is the use of the correct and complete D3-brane action. In particular, a key $B_2$-$C_2$ coupling term on D3-branes, which surprisingly is missing in standard textbooks \cite{Polchinski:1998rr,Becker:2006dvp,Blumenhagen:2013fgp}, must be included to find the zero result. We devote 
App.~\ref{app:ExtraB2C2term} to explaining the origin of this term in detail.
In sect.~\ref{sc:GCwF}, we include $H_3/F_3$ fluxes, breaking  SL$(2,\mathbb{R})$ and inducing different masses for $B_2$ and $C_2$ as well as a mixing with $C_4$. This generically destroys the cancellation and leads to our desired non-zero mixing result, which is however suppressed in the large volume limit by both sequestering and by the diluteness of the 3-form flux. Since our calculation implements the SL$(2,\mathbb{R})$ structure and hence electric-magnetic duality on D3-branes, we also obtain an explicit expression for the magnetic mixing. 
This complements magnetic mixing results in the non-sequestered strong-coupling regime discussed in \cite{Heckman:2011sw,DelZotto:2016fju}.
We use our findings to derive some first phenomenological implications in sect.~\ref{sc:pheno}. We conclude in sect.~\ref{sc:conclusion}. 

\subsection{Phenomenological motivation}
\label{pmot}
The phenomenology of hidden photons is a very rich subject. We briefly highlight certain aspects, which we deem interesting for the following stringy scenarios and motivate the topic.

The hidden U(1) may be massless or massive. In the second case, the canonical kinetic terms and \eqref{eq:kinetic-mixing-def} are supplemented by
\be
\label{eq:massive}
 	\mathcal{L} \supset -\frac{m^2}{2} \mathbb{A}_\fieldlable{(B)}^{\mu} \mathbb{A}_\mu^\fieldlable{(B)}~.
\ee
Most simply, such a term can originate from a Higgs or a Stueckelberg mechanism using a field charged only under 
U$(1)_\fieldlable{(B)}$, such that 
the visible U$(1)_\fieldlable{(A)}$ remains massless.
If massive, the dark photon could itself play the role of dark matter~\cite{Nelson:2011sf,Arias:2012az,Graham:2015rva,Agrawal:2018vin,Co:2018lka}, otherwise further fields need to be included to serve that purpose. We also note that, due to KM, dark photons are a messenger to dark matter~\cite{Arkani-Hamed:2008hhe} and more generally to the hidden sector.

The visible U(1)$_\fieldlable{(A)}$ obviously couples to the current $j^\mu_\fieldlable{(A)}$ containing the charged Standard Model states, which are relatively light.
For KM to be observable, one requires that U(1)$_\fieldlable{(B)}$ either also possesses light charged states,
\be
 {\cal L} \supset g_\fieldlable{(A)}~j^\mu_\fieldlable{(A)} \mathbb{A}_\mu^\fieldlable{(A)} + g_\fieldlable{(B)}~j^\mu_\fieldlable{(B)} \mathbb{A}_\mu^\fieldlable{(B)}~,
\ee
or that it is massive, cf.~\eqref{eq:massive}.
Otherwise, all effects of KM can be removed by a field redefinition. 

If light states charged under the hidden U(1)$_\fieldlable{(B)}$ are present, then the best one can achieve is a diagonal field basis in which the SM states are not charged under U(1)$_\fieldlable{(B)}$ but the hidden-sector states have a small charge $Q$ under U(1)$_\fieldlable{(A)}$,
\begin{equation}\label{eq:mcp-def}
    {\cal L} \supset Q ~g_\fieldlable{(A)}~j_\fieldlable{(B)}^\mu~\mathbb{A}_\mu^\fieldlable{(A)}~,\quad Q=\kmp_\fieldlable{AB}\frac{g_\fieldlable{(B)}}{g_\fieldlable{(A)}}\,.
\end{equation}
These ``millicharged'' states represent an experimental target. 
If, by contrast, the hidden gauge boson $\mathbb{A}_\mu^\fieldlable{(B)}$ is massive, then one can find a basis with diagonal mass and kinetic term where the SM states are millicharged under U(1)$_\fieldlable{(B)}$. Now the experimental target is represented by a massive gauge boson with tiny coupling to SM particles.

\begin{figure}[!t]
    \centering
    \includegraphics[width=0.96\textwidth]{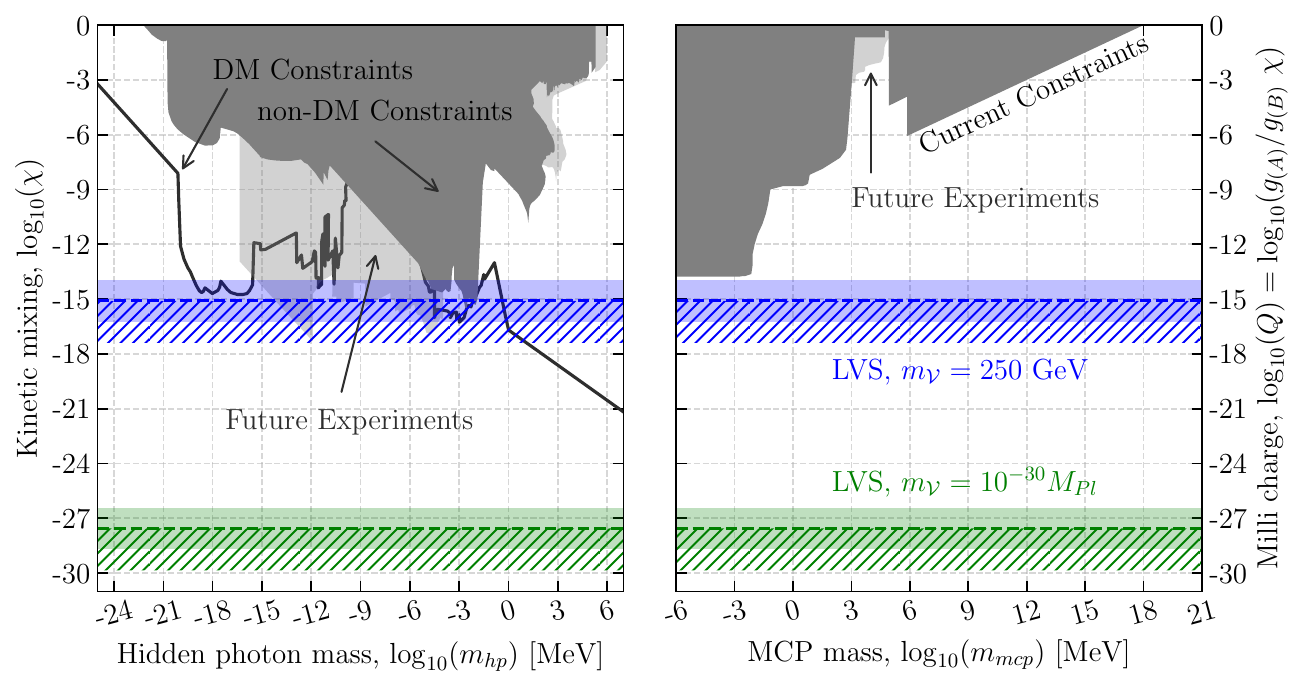}
    \caption{
    Current constraints on KM in case of a massive hidden photon (left) and a massless hidden photon with millicharged particles (MCP) (right); adapted mainly from~\cite{AxionLimits,Chang:2016ntp,Fabbrichesi:2020wbt,Araki:2023xgb,Goodsell:2009pi,Jaeckel:2012yz}. 
    The colored delimiters show the indicative lower values obtained for D3-D3 brane setups with fluxes, as discussed in sect.~\ref{sc:pheno}. See text for details and further references.
    }
    \label{fig:KM-Parameter-Plots}
\end{figure}

The resulting different sets of constraints relevant in the massless and massive case are illustrated in fig.~\ref{fig:KM-Parameter-Plots}.

 The left panel of fig.~\ref{fig:KM-Parameter-Plots} shows the case of massive hidden photons. We distinguish between direct constraints relying solely on the existence of the hidden photon (dark gray region) and constraints assuming the hidden photon is dark matter (solid line). The data is combined and adapted from the collections~\cite{AxionLimits,Chang:2016ntp,Fabbrichesi:2020wbt,Araki:2023xgb} where also the original references can be found. 
The DM constraint is extended to masses larger than an MeV accounting for decays into electron-positron pairs and requiring that the lifetime is greater than the age of the Universe. This is analogous to what was done for photons, e.g., in~\cite{Arias:2012az} and using decay formulae from~\cite{Pospelov:2008zw,Andreas:2012mt}.
The light gray region is to be covered by future experiments, see refs. in~\cite{AxionLimits}.

The right panel of fig.~\ref{fig:KM-Parameter-Plots} shows the case of massless hidden photons with massive particles charged under it (see again~\cite{Goodsell:2009pi,Jaeckel:2012yz,Fabbrichesi:2020wbt} and refs. therein). The gauge couplings are assumed to be equal, such that $Q=\kmp$. As seen in the figure, a region of high millicharge $Q$ is excluded by CMB limits~\cite{Dubovsky:2003yn} and the overabundance of DM~\cite{Davidson:1991si} (adapted to the case where $e=g_\fieldlable{(A)}=g_\fieldlable{(B)}$).
Again, the dark gray region refers to current constraints and the light gray region to future experiments. 
For a more general and detailed discussion of the constraints see e.g.~\cite{Jaeckel:2010ni,Jaeckel:2012mjv,Fabbrichesi:2020wbt,AxionLimits}.

Encouragingly, the setup studied in this paper, though admittedly oversimplified and not realistic, provides values that are in regions of phenomenological interest and potentially probed in future experiments.

\section{SL\boldmath{$(2,\mathbb{R})$} structure}\label{sc:SL2R-notations}
We consider type IIB supergravity with D3-branes and start by clarifying its SL$(2,\mathbb{R})$ symmetry. The 10d bulk action reads
\cite{Bergshoeff:1995as,Bergshoeff:1995sq}\footnote{\label{fn:epsilon}We chose the conventions $\epsilon^{12}=\epsilon_{21}=1$ and $\epsilon_{12}=\epsilon^{21}=-1$.}
\begin{equation}
\begin{aligned}
\label{eq:IIBaction}
S_{\text{IIB}} = \frac{1}{2\kappa_\text{10}^2}\int\limits_{\M^{10}} \dx{10}{x} &\sqrt{-G_E}~ \left( R_E - \frac{\partial_N \Bar{\tau} \partial^N \tau}{2 (\Im ~\tau)^2} \right) \\
&+\frac{1}{2\kappa_\text{10}^2}\int\limits_{\M^{10}} \left( -\frac{\hat{M}_{ij}}{2} F^i_3 \whs{10} F^j_3 -\frac{1}{4} \tilde{F}_5 \whs{10} \tilde{F}_5 - \frac{\epsilon_{ij}}{4} C_4\wedge F^i_3 \wedge F^j_3 \right)~,
\end{aligned}
\end{equation}
with $N$ a 10d index and $2\kappa_{10}^2=(2\pi)^7\ap^4$. The indices $i$, $j$ label the two 3-form field strengths
\begin{equation}
    F_3^{i}=\d C^i_2=\begin{pmatrix}
        \d C_2\\\d B_2
    \end{pmatrix}~.
\end{equation}
We also define
\begin{gather} 
    \label{eq:tauDef}
    \tau = C_0 + i e^{-\phi}~,\\
    \label{eq:Def-M}
    \hat{M}_{ij}= \frac{1}{\Im ~\tau} \begin{pmatrix} 1 & -\Re ~\tau \\ -\Re ~\tau & |\tau|^2  \end{pmatrix}~,\\
    \tilde{F}_5 = \d C_4 + \frac{\epsilon_{ij}}{2}  C_2^i\wedge F_3^j ~.
\end{gather}
The action \eqref{eq:IIBaction} is invariant under the gauge transformations
\begin{gather}
    \label{eq:C2B2-gaugetrafo}
    \delta C_2^i = \d \Lambda_1^i~,
    \\
    \label{eq:C4gaugetrafo}
    \delta C_4 = \d\Lambda_3 - \frac{\epsilon_{ij}}{2}  \Lambda_1^i\wedge F_3^j
\end{gather}
as well as under the global SL$(2,\mathbb{R})$ transformations
\begin{align}
    \label{eq:tau-Trafo}
    \tau' &= \frac{a\tau + b}{c\tau +d}~, &\Lambda^i_{~j}&=\begin{pmatrix}
    a&b\\c&d
    \end{pmatrix}~,\\
    \label{eq:C2B2SL2RTrafo}
    C'^i_2
    &= \Lambda^i_{~j} C_2^j~,  &\hat{M}' &= (\Lambda^{-1})^T \hat{M} \Lambda^{-1}\,.
\end{align}
The form $C_4$ and the Einstein-frame metric $G_E$ do not transform.

To quadratic order in the gauge field strength $F_2$, the Einstein-frame action for a D3-brane reads \cite{Fradkin:1985qd,Abouelsaood:1986gd,Dai:1989ua,Leigh:1989jq,Witten:1995im,Polchinski:1995mt,Li:1995pq,Douglas:1995bn,Bershadsky:1995qy,Green:1996bh,Green:1996dd,Witten:1996hc,Mourad:1997uc,Cheung:1997az,Minasian:1997mm}
\begin{align}
    \label{eq:D3-action}
    S_{D3} = &~S_\actionlabel{DBI} + S_\actionlabel{WZ}~,\\
    \label{eq:DBI-action}
     S_\actionlabel{DBI}= &-\bt_3 \int\limits_{D3}   \frac{e^{-\phi}}{2}\left(F_2 -B_2 \right) \whs{4} \left(F_2 -B_2 \right)  ~,
     \\
     \label{eq:D3-CS-action}
     S_\actionlabel{WZ}=&~
     \bt_3 \int\limits_{D 3} C_4 +\frac12 B_2\wedge C_2 + C_2 \wedge \left(F_2 -B_2 \right) + \frac{C_0}{2}\left(F_2 -B_2 \right) \wedge \left(F_2 -B_2 \right)
 	\,,
\end{align}
with the brane tension given by $\bt_p=2\pi~(2\pi \sqrt{\ap})^{-(p+1)}$. While, for our purposes, this action is simply dictated by string theory, we want to highlight a recent discussion on generalizations of similar actions in field theory in the context of axion physics \cite{Sokolov:2022fvs, Heidenreich:2023pbi}.

We emphasise that it will be crucial for us to work with the correct and complete D3-brane action. As noted in the introduction, the second term on the r.h.s. of \eqref{eq:D3-CS-action} is missing in several standard textbooks \cite{Polchinski:1998rr,Becker:2006dvp,Blumenhagen:2013fgp}. However, this $B_2$-$C_2$ coupling is essential for gauge invariance and SL$(2,\mathbb{R})$ self-duality of the D3-brane. It can be found by going back to the original literature \cite{Bergshoeff:1995as,Bergshoeff:1995sq,Green:1996bh,Ortin:2015hya,Douglas:1995bn,Morrison:1995yi,Tseytlin:1996it,Bergshoeff:1996cy,Kimura:1999jb,Bergshoeff:2006gs}.
We comment on this term in more detail in  app.~\ref{app:ExtraB2C2term}.

The brane action \eqref{eq:D3-action} is invariant under the gauge transformations \eqref{eq:C2B2-gaugetrafo} and \eqref{eq:C4gaugetrafo} together with the transformation of the gauge field strength
\begin{equation}
    \label{eq:F2-gaugetrafo}
    \delta F_2 = d \Lambda_1^{(2)}~.
\end{equation}
To study SL$(2,\mathbb{R})$ transformation properties, we need to specify the transformation of the gauge field strength $F_2$.
Similarly to $B_2$, the field strength $F_2$ transforms as part of a doublet which it forms together with its dual field strength $G_2$ \cite{Tseytlin:1996it,Green:1996bh},
\begin{equation}
    \label{eq:F2-SL-trafo}
    \begin{pmatrix}
        G'_2\\F'_2
    \end{pmatrix}
    = \begin{pmatrix}
    a&b\\c&d
    \end{pmatrix}\begin{pmatrix}
        G_2\\F_2
    \end{pmatrix}~.
\end{equation}
 Formally, the dual field strength $G_2$ is defined by \cite{Gaillard:1981rj,Gaillard:1997rt}\footnote{We use a different sign convention for $G_2$ compared to \cite{Gaillard:1981rj,Gaillard:1997rt}.}
\begin{equation}
    \delta {\cal{L}} = \delta F_2 \wedge G_2~,
\end{equation}
which for the D3-brane explicitly yields
\begin{equation}
    G_2 = - e^{-\phi} \hs{4} (F_2-B_2) + C_0 (F_2-B_2) + C_2~.
\end{equation}

Contrary to widespread beliefs, the D3-brane action is not SL$(2,\mathbb{R})$ invariant. This can be seen by performing an infinitesimal SL$(2,\mathbb{R})$ transformation defined by
\begin{equation}
    \delta\Lambda^i{}_j = \begin{pmatrix}
        \alpha & \beta\\ \gamma & -\alpha
    \end{pmatrix}~.
\end{equation}
It produces the following change of the D3-brane action,
\begin{equation}\label{eq:infi-sl-trafo-of-action}
    \delta_{SL} S_{D3} = \int\limits_{D3} \frac{\beta}{2} F_2 \wedge F_2 + \frac{\gamma}{2} G_2 \wedge G_2~,
\end{equation}
where the second term is \textit{not} a total derivative.
This is in agreement with the general treatment of self-duality in \cite{Gaillard:1981rj,Gaillard:1997rt}:
Although the action is not invariant, the equations of motion of bulk and brane fields are invariant under SL$(2,\mathbb{R})$ transformations, i.e. the theory is SL$(2,\mathbb{R})$ self-dual in the spirit of \cite{Gaillard:1981rj,Gaillard:1997rt} which was recently emphasized in~\cite{Russo:2024llm}.
On the other hand, when restricting ourselves to SL$(2,\mathbb{Z})$ transformations generated by
\begin{equation}
    \cal{I} = \begin{pmatrix} 0 &-1\\1&0\end{pmatrix}~,\quad \cal{T}=\begin{pmatrix} 1 &1\\0&1\end{pmatrix}~,
\end{equation}
the action is in fact invariant under $\cal T$ and mapped to its dual theory by $\cal I$. This mapping is a specific instance \cite{Green:1996bh,Tseytlin:1996it} of the general concept \cite{Gaillard:1981rj,Gaillard:1997rt} of mapping to a dual theory.
It produces a dual Lagrangian ${\cal L}_{dual}={\cal L}_{dual}(G_2,\tilde{g})$, where $\tilde{g}$ is the dual coupling. In our case, the theory is said to be self-dual because the relation
\begin{equation}
    {\cal{L}}_{dual}(G_2,\tilde{g})\equiv {\cal{L}}(F_2=G_2,g=\tilde{g})
\end{equation}
holds.

Although SL$(2,\mathbb{R})$ is not a symmetry, it is convenient to rewrite the D3-brane action \eqref{eq:D3-action} using SL$(2,\mathbb{R})$ indices, in a manner similar to \eqref{eq:IIBaction}. At quadratic order we have
\begin{equation}
    \label{eq:D3-action-rewritten}
    S_{D3} = \bt_3 \int\limits_{D3} C_4 +\frac12 J_{(1)}\wedge J_{(2)} -\frac12 C^i_2 \whs{4} \hat{m}_{ij} C^j_2 + C^i_2\whs4 J_i~,
\end{equation}
where we defined
\begin{gather}
    \label{eq:F-sources}
    J_{(1)} = -\hs{4} F_2~, \quad J_{(2)}= \gs^{-1} F_2 + C_0 \hs{4} F_2~,\\
    \label{eq:self-coupling-matrix}
    \hat{m}_{ij}=\begin{pmatrix}
        0 & -\frac12 \hs{4}\\ -\frac12 \hs{4} & \gs^{-1} + C_0 \hs{4}
    \end{pmatrix}~.
\end{gather}
Note that $J_i$ and $\hat{m}_{ij}$ do not transform as a vector and tensor under the full  SL$(2,\mathbb{R})$, but only under the (Borel) subgroup $\mathfrak{b}$ generated by
\begin{equation}
    \label{eq:SL-Subgroup}
    \delta\Lambda^i{}_j = \begin{pmatrix}
        \alpha & \beta\\ 0 & -\alpha
    \end{pmatrix}~.
\end{equation}
This is a proper, 2-parameter symmetry group of $S_{D3}$, cf. \eqref{eq:infi-sl-trafo-of-action}.

\section{Leading order cancellation}\label{sc:LOC}
To begin we repeat the calculation of~\cite{Abel:2008ai}, showing that kinetic mixing vanishes in the D3 scenario at leading order and without fluxes.
We consider a generic CY-orientifold with O3/O7-planes, thereby projecting out the massless modes of $B_2$ and $C_2$. This allows us to integrate out these fields by solving their equation of motion.

Following \cite{Abel:2008ai}, we set  $C_0=C_4=0$ and consider constant $e^\phi=\gs$. In doing so, the matrix $\hat{M}$ from \eqref{eq:Def-M} becomes diagonal. This implies that there is no mixing between $C_2$ and $B_2$ in the bulk action \eqref{eq:IIBaction}. Moreover, we neglect $\hat{m}$, i.e. all self couplings of $B_2$ and $C_2$ on the D3-branes.
This reduces the relevant parts in the actions \eqref{eq:IIBaction} and \eqref{eq:D3-action-rewritten} to
\begin{equation}
    \label{eq:action-sect-LOcancellation}
    S = \frac{1}{2\kappa_\text{10}^2}\int\limits_{\M^{10}}  -\frac{\hat{M}_{ij}}{2} F^i_3 \whs{10} F^j_3 + \bt_3\int\limits_{\M^{10}} C_2^i \wedge \left[\hs{4}J_i\wedge \delta_6(A)+\hs{4}J_i\wedge\delta_6(B)\right]~.
\end{equation}
Here $\delta_6(A/B)=\delta_6(y-y_{A/B})$ are 6-form delta functions localized at the positions of brane $A$ and brane $B$ in the compact space parameterized by $y$.
Furthermore, the sources $J_i$ simplify to
\begin{equation}
    J_{(1)} = -\hs{4} F_2~,\quad J_{(2)}=\gs^{-1}F_2~.
\end{equation}
The pairing of $J_i$ with the delta distributions fixes whether we have to use $F_2^{(A)}$ or $F_2^{(B)}$ in this definition. In order to not clutter the notation, we therefore did not give $J_i$ labels $A/B$ and will also drop these labels on $F_2$ for now. We will restore them whenever we think it is useful for clarity.  The equations of motion for $C^i_2$ following from \eqref{eq:action-sect-LOcancellation} now read
\begin{equation}
    \label{eq:eomCi}
    \codiff \d C^i_2 = 2\kappa_\text{10}^2\bt_3 \left((\hat{M}^{-1})^{ij} J_j ~\delta(A) +(\hat{M}^{-1})^{ij} J_j~\delta(B)\right)~,
\end{equation}
with scalar distributions for the brane positions $\delta(A/B)=\hs{6}\delta_6(A/B)$.
For the sake of clarity, we will absorb the factor $2\kappa_\text{10}^2\bt_3$ into the six-dimensional $\delta$-function and Laplace operator, respectively:  
\begin{equation}\label{eq:factor-absorption}
    2\kappa_\text{10}^2\bt_3~ \delta(y-y_{A/B}) \eqqcolon \pair{\delta}(y-y_{A/B})~,\quad 2\kappa_\text{10}^2\bt_3~\Delta^{-1}_6 \eqqcolon \pair{\Delta}^{-1}_6~,
\end{equation}
This defines underlined quantities. Again, we will reintroduce the factor $2\kappa_\text{10}^2\bt_3$ only when helpful.
It is obvious from the above equation of motion that only $B_2$ and $C_2$ fields with 4d indices are sourced. Thus we treat the 2-forms $C^i_2$ as scalars on the 6d internal manifold $\intmfd$. 
Fixing the gauge by
\begin{equation}
    \codiff C^i_2=0~,
\end{equation}
simplifies
\begin{equation}
    \Box_{10} \equiv \codiff\dx{}{} + \dx{}{}~\codiff = \codiff\dx{}{}~,
\end{equation}
when acting on $C^i_2$. Since the length scales on which our sources vary in 4d are much larger than the size of the compact space, 
we may neglect 4d derivatives w.r.t. 6d derivatives. This implies 
\begin{equation}
    \Box_{10}C^i_2 = \Delta_6C^i_2~.
\end{equation}

\begin{figure}[t]
    \centering
    \subfloat[]{\includegraphics[height=2.5cm,valign=c]{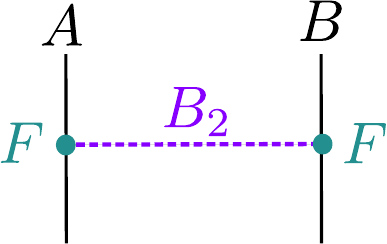}
    }
    \hspace{13pt}
    \subfloat[]{\includegraphics[height=2.5cm,valign=c]{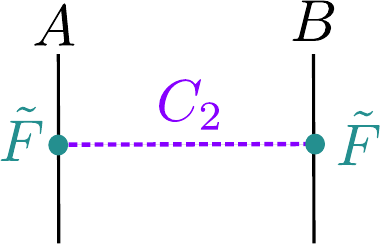}
    }
    \caption{Leading-order diagrams relevant for KM: $B_2$ exchange (a) and $C_2$ exchange (b).}
    \label{fig:LO-diagrams}
\end{figure}

Under this assumption, the solutions to \eqref{eq:eomCi} are simply given in terms of the Green's function\footnote{Note that the basic source term $F_2$ relevant in our context is odd under the orientifolding. Hence the integrated source vanishes in the `upstairs' geometry before orientifolding, making the compact-space Green's function well-defined.} $\Delta_6^{-1}$ on the internal manifold $\intmfd$
\begin{equation}
    C^i_2(x,y)= - \pair{\Delta}_6^{-1}(y,y_A) (\hat{M}^{-1})^{ij} J_j(x) - \pair{\Delta}_6^{-1}(y,y_B) (\hat{M}^{-1})^{ij} J_j(x)~,
\end{equation}
where $x$ and $y$ refer to 4d and 6d coordinates respectively. Replacing $C^i_2$ in \eqref{eq:action-sect-LOcancellation} with this solution, i.e. integrating out $C_2^i$, we obtain an effective action which contains the mixing terms
\begin{equation}
    \label{eq:result1}
    S\supset \frac{\bt_3}{2} \int\limits_{\extmfd}\pair{\Delta}_6^{-1}(y_A,y_B)  \left[J^{(A)}_i  \whs{4}~ (\hat{M}^{-1})^{ij} J^{(B)}_j + J^{(B)}_i  \whs{4}~ (\hat{M}^{-1})^{ij} J^{(A)}_j\right] ~.
\end{equation}
Here we reinstated the brane labels and made use of the symmetry of the Green's function, which can always be imposed on compact Riemannian manifolds.
Now one observes that
\begin{equation}
    \label{meps}
    (\hat{M}^{-1})^{ij} J_j(x) = - \hs{4} \epsilon^{ij} J_j~,
\end{equation}
as can be checked explicitly.
This relation implies that the square bracket in \eqref{eq:result1} takes the form
\begin{equation}
    \label{eq:cancellation}
    \epsilon^{ij} \left(J^{(A)}_i  \wedge J^{(B)}_j +  J^{(B)}_i  \wedge  J^{(A)}_j\right) = 0\,,
\end{equation}
vanishing by (anti-)symmetry of the two factors. Thus, the findings of
\cite{Abel:2008ai} on tori and of
\cite{Dienes:1996zr,Abel:2003ue} on Calabi-Yaus are reproduced. However, this result hinges on the simplifications made at the beginning of this section by setting $C_0=0$ and disregarding brane-localized mixing terms between $B_2$ and $C_2$. 
When we drop these simplifications in the next section, it will turn out to be useful that we formulated our analysis in such a way that the crucial cancellation follows from the SL$(2,\mathbb{R})$ index structure. We note for completeness that one may equivalently ascribe the zero result to a cancellation between the effects of $B_2$ and $C_2$ exchange between the branes, cf. fig. \ref{fig:LO-diagrams}. 
The exchange of $B_2$ gives a term proportional to $F^\fieldlable{(A)}_{\mu\nu}F^{\mu\nu}_\fieldlable{(B)}$, while the exchange of $C_2$ gives $\tilde{F}^\fieldlable{(A)}_{\mu\nu}\tilde{F}^{\mu\nu}_\fieldlable{(B)}=-F^\fieldlable{(A)}_{\mu\nu}F^{\mu\nu}_\fieldlable{(B)}$. 

\section{General cancellation without fluxes}\label{sc:GCwoF}
In flux compactifications, $C_0$ is generically stabilized at a non-zero ${\cal O}(1)$ value \cite{Giddings:2001yu}. This will introduce mixing of $B_2$ and $C_2$ through the matrix $\hat{M}$ -- a leading-order effect which has to be taken into account. Moreover, there exist brane-localized mixing terms between $B_2$ and $C_2$, encoded in the matrix $\hat{m}$. As we will discuss in app.~\ref{app:issues}, they lead to UV-divergent diagrams, such that these mixing terms are a priori important.

Therefore, we will now extend the analysis of sect.~\ref{sc:LOC} (and hence of~\cite{Abel:2008ai}) by including these generalizations.
All we have to do is to repeat the calculation in sect.~\ref{sc:LOC} with the complete expression for $\hat{M}$ and with the full D3-brane action,
\begin{equation}
    S_{D3}=\bt_3 \int\limits_{D3} C^i_2 \whs{4} J_i - \frac12 C^i_2 \whs{4} ~\hat{m}_{ij} C^j_2  ~.
\end{equation}
Here the sources $J_i$ are now given by \eqref{eq:F-sources}, including $C_0$, and we recall that the self-coupling matrix $\hat{m}_{ij}$ can be found in \eqref{eq:self-coupling-matrix}.
Applying these generalizations, the equations of motion for $C_2^i$ read\footnote{\label{fn:dwch} 
We 
will see in sect.~\ref{sc:feom} that when carefully accounting for, so far neglected, $C_{4}$ and $\tilde{F}_{5}$ terms, the equation of motion for $C^{i}_{2}$ acquires an extra contribution. This can be seen from \eqref{eq:fluxeom} when recalling $\hat{M}_{ij}(\codiff \d)  ~C_2^j=\hs{10}\d \left( \hat{M}_{ij} \hs{10} F_3^j \right)$. The extra term may be easily accounted for by replacing $\hat{m}$ with $\tilde{m}$, given by \eqref{eq:mtilde}. The key features of $\hat{m}$ and hence the result of this section remain unchanged.
}
\begin{align}
    \label{eq:eom2}
    \hat{D}_{ij}(\codiff \d)  ~C_2^j = J_i ~\pair{\delta}(A) + J_i ~\pair{\delta}(B)
\end{align}
where
\begin{align}
    \hat{D}_{ij}(\codiff \d) \equiv  \left[\hat{M}_{ij} ~\codiff \d  + \hat{m}_{ij} ~\pair{\delta}(A) + \hat{m}_{ij}~ \pair{\delta}(B)\right]~,
\end{align}
and we absorbed a factor of $2 \kappa^2_{10} \bt_3$ in the $\delta$-functions, cf.~\eqref{eq:factor-absorption}.
To integrate out $C_2^i$ we fix the gauge and neglect 4d fluctuations analogously to sect.~\ref{sc:LOC}.

\begin{figure}[t]
    \centering
    \subfloat[]{\includegraphics[height=2.5cm,valign=c]{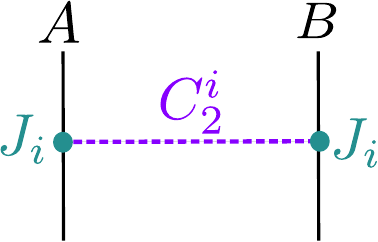}
    }
    \hspace{13pt}
    \subfloat[]{\includegraphics[height=2.5cm,valign=c]{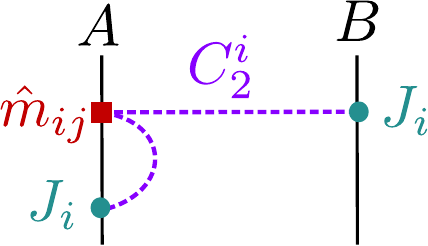}
    }
    \hspace{13pt}
    \subfloat[]{\includegraphics[height=2.5cm,valign=c]{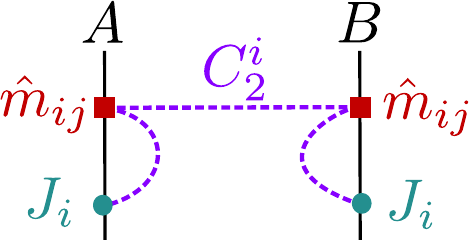}
    }
    \caption{Due to the new coupling $\hat{m}_{ij}$ on each D3-brane, there are now more diagrams contributing to KM. Figures~(a), (b) and (c) provide example diagrams in perturbation theory at different orders. As we show, all contributions vanish individually to all orders in perturbation theory.}
    \label{fig:m-diagrams}
\end{figure}

Thus we are left with the problem of inverting the operator $\hat{D}(\Delta_6)$ which we solve using a series expansion in $\hat{m}$:
\begin{equation}
    \hat{D}^{-1}(\Delta_6) 
    = \sum\limits_{k=0}^\infty \hat{M}^{-1}\Delta_6^{-1} \left(\left[ \hat{m} ~\pair{\delta}(A) + \hat{m}~ \pair{\delta}(B) \right] \hat{M}^{-1}\Delta_6^{-1}\right)^k~.
\end{equation}
This allows us to integrate out $C^i_2$. One may think of the effects we calculate in terms 
of diagrams describing $C_2^i$-exchange between branes, as depicted in fig.~\ref{fig:m-diagrams}. The resulting action is analogous to \eqref{eq:result1} and reads
\begin{equation}
    \label{eq:result2}    S\supset \frac{\bt_3}{2} \int\limits_{\extmfd}  \pair{J}^{(A)}_i \whs{4} ~(\hat{D}^{-1})^{ij}~ J^{(B)}_j +\pair{J}^{(B)}_i \whs{4} ~(\hat{D}^{-1})^{ij}~ J^{(A)}_j~.
\end{equation}
There are two important observations to be made,
\begin{gather}
    (\hat{M}^{-1})^{ij} J_j(x) = - \hs{4} \epsilon^{ij} J_j\equiv  - \hs{4} J^i~,\\
    \label{eq:mhatrelation}
    \hat{m}_{ij} \hs{4}J^j(x) = -\frac{1}{2} \epsilon_{ij} J^j \equiv -\frac{1}{2} J_i~,
\end{gather}
which can be checked by explicit calculation. The first simply generalizes \eqref{meps} to $C_0\neq 0$. The second is an analogous relation for $\hat{m}$. We have also introduced the natural notation $J^i\equiv \epsilon^{ij}J_j$ \footref{fn:epsilon}.

As a result, the application of $\hat{D}^{-1}$ to the source now gives\footnote{As already remarked in footnote~\ref{fn:dwch} $\hat{m}$ should actually be replaced by $\tilde{m}$. This changes \eqref{eq:mhatrelation} into \eqref{eq:mtilderelation}. While the pre-factor is now $1$ instead of $1/2$, the decisive property of lowering the index $i$ on $J^{i}$ persists.  }
\begin{equation}\label{eq:inverted-propagator}
    (\hat{D}^{-1})^{ij} ~J_j = - \sum\limits_{k=0}^\infty \Delta_6^{-1} \left(\frac{\Delta_6^{-1}}{2} ~\pair{\delta}(A) + \frac{\Delta_6^{-1}}{2}~ \pair{\delta}(B) \right)^k~\hs{4}J^i~.
\end{equation}
When we now evaluate \eqref{eq:result2} using this result, we obtain an infinite sum of terms each of which is proportional to
\begin{equation}
    \label{eq:cancellation1}
    \epsilon^{ij} \left(J^{(A)}_i  \wedge J^{(B)}_j +  J^{(B)}_i  \wedge  J^{(A)}_j\right) = 0\,.
\end{equation}
We conclude that kinetic mixing between D$3$-branes vanishes in full generality.

The structure of the cancellations suggests that the underlying reason is the very peculiar SL$(2,\mathbb{R})$ structure of D3-branes in type IIB.
To be more precise, even though SL$(2,\mathbb{R})$ is not a symmetry of the action, the subgroup $\mathfrak{b}$ is a symmetry and the sources $J_i$ are doublets under it. Moreover, it is easy to check explicitly that, just like in the case of SL$(2,\mathbb{R})$, the matrix $\epsilon^{ij}$ is the only rank-2 invariant tensor of the group $\mathfrak{b}$.
At the lowest non-trivial order contributing to the kinetic mixing we have to build an invariant from the two $\mathfrak{b}$-vectors, $J_i$. Contraction with $\epsilon^{ij}$ then provides the unique $\mathfrak{b}$-invariant (and SL$(2,\mathbb{R})$-invariant) way to combine the two sources.
Therefore, the $\mathfrak{b}$ symmetry together with the obvious symmetry under exchange of $A/B$ ensures that our final result will be proportional to the l.h.s.~of \eqref{eq:cancellation1} and hence vanish. The calculation above can be regarded as an explicit confirmation.

\section{No cancellation including fluxes}\label{sc:GCwF}
\subsection{Deriving relevant equation of motion}\label{sc:feom}

In a final step we now include a general internal background flux $\bcgr{F}_3^i$ for the field strength $F_3^i$, breaking the SL$(2,\mathbb{R})$ symmetry. 
Having $\bcgr{F}_3^i$ fluxes at our disposal, more terms quadratic in the sources $J_{i}$ can be constructed. Due to Lorentz invariance, at least two fluxes $\bcgr{F}_3^i$ have to be used, with $J_{i\,\mu\nu} J_{j}^{\mu\nu} \bcgr{F}_{abc}^i \bcgr{F}^{j\, abc}$ being the minimal option. Otherwise Lorentz indices would be left open. Indeed, our explicit result \eqref{eq:final-KM} has this form.

To begin we note that, after compactification, the $\Tilde{F}_5\whs{10}\Tilde{F}_5$ term in the action \eqref{eq:IIBaction} induces masses for $C_2^i$ \cite{Abel:2008ai}. In the analysis of \cite{Abel:2008ai}, the relevant mass terms explicitly contain the background gauge fields $\bar{C}_2^i$. We find it problematic that the latter are not gauge invariant and can not be globally defined on the compact space. We were not able to find a gauge-invariant rewriting at the level of the action. The equations of motion however are fully gauge invariant and thus allow for a consistent analysis. To obtain the equations of motion we recall the relevant parts of the bulk action
\begin{gather}
    S_{IIB}\supset\frac{1}{2\kappa_\text{10}^2}\int\limits_{\M} \left( -\frac{\hat{M}_{ij}}{2} F^i_3 \whs{} F^j_3 -\frac{1}{4} \tilde{F}_5 \whs{} \tilde{F}_5 - \frac{\epsilon_{ij}}{4} C_4\wedge F^i_3 \wedge F^j_3 \right)~,\\
    \label{eq:def-tilde-F5}
    \tilde{F}_5= \d C_4 + \frac{\epsilon_{ij}}{2} C_2^i \wedge F_3^j
\end{gather}
and the D3-brane action
\begin{equation}
    S_{D3} =  \int\limits_{\M}\bt_3\left( C_4 +\frac12 J_{(1)}\wedge J_{(2)} -\frac12 C^i_2 \whs{4} \hat{m}_{ij} C^j_2 + C^i_2\wedge J_i~\right)\wedge\delta_6(y-y_{D3})~.
\end{equation}
It is convenient to first consider the equation of motion for $C_4$. Since a D3-brane acts as an electric and magnetic source for $C_4$, the correct equation of motion is obtained by varying the action w.r.t. $C_4$ while considering only half of the D3-brane contribution.\footnote{Cf. footnote 6 of \cite{Giddings:2001yu}.} This gives
\begin{equation}
    \label{eq:eomC4}
    \d\hs{10} \tilde{F}_5 = \frac{\epsilon_{ij}}{2} F_3^i \wedge F_3^j - \pair{\delta}_6(y-y_{D3})~,
\end{equation}
which as usual is further constrained by imposing self-duality, $\tilde{F}_5=\hs{10}\tilde{F}_5$. Again, we employ our short-hand notation introduced in \eqref{eq:factor-absorption}.

Varying the action w.r.t. $C_2^i$ yields the equation of motion for $C_2^i$ \footnote{Here we used the identities $\hat{m}_{ij}=\hat{m}_{ji}$ and $A_2^i\whs{4}\hat{m}_{ij}B_2^j= B_2^i\whs{4}\hat{m}_{ij}A_2^j$ for arbitrary $A_2^i$ and $B_2^j$.},
\begin{equation}
\begin{aligned}
    \d \left( \hat{M}_{ij} \hs{10} F_3^j \right) 
    = \frac{\epsilon_{ij}}{2}  \left( F_3^j\whs{10} \Tilde{F}_5 + F_3^j \wedge \d C_4 \right)
    + \frac{\epsilon_{ij}}{4} &C_2^j\wedge \d \hs{10} \Tilde{F}_5 \\
     &- \left(\hs{4} J_i - \hs{4} \hat{m}_{ij} C_2^j\right) \wedge \pair{\delta}_6~.
\end{aligned}
\end{equation}
This equation can now be rewritten by using the definition \eqref{eq:def-tilde-F5} and the equation of motion for $C_4$ \eqref{eq:eomC4}:
\begin{equation}
\label{eq:fluxeom}
    \d \left( \hat{M}_{ij} \hs{10} F_3^j \right) 
    =  \epsilon_{ij} F_3^j \wedge \tilde{F}_5- \left(\hs{4} J_i - \hs{4} \hat{m}_{ij} C_2^j +\frac{\epsilon_{ij}}{2} C_2^j\right) \wedge \pair{\delta}_6~.
\end{equation}
One may check explicitly that the two terms containing $C_2^j$ on the r.h. side of \eqref{eq:fluxeom} combine with the brane field $F_2$ contained in $J_i$ to form a gauge invariant expression:
\begin{equation}
    \hs{4} J_i[F_2] - \hs{4} \hat{m}_{ij} C_2^j +\frac{\epsilon_{ij}}{2} C_2^j= \hs{4} J_i[F_2-B_2] ~.
\end{equation}
Thus, our equation of motion is gauge invariant. To simplify it further, we define a new self-coupling matrix $\tilde{m}$ by
\begin{equation}
\label{eq:mtilde}
    \hs{4}\tilde{m}_{ij}\equiv\hs{4} \hat{m}_{ij} -\frac{\epsilon_{ij}}{2} =\hs{4}\begin{pmatrix}0&-\hs{4}\\0&\gs^{-1} + C_0 \hs{4}
    \end{pmatrix} \,. 
\end{equation}
Note that the second equality holds only when this matrix is applied to a 2-form. As a key feature, our new $\tilde{m}$ still possesses a property analogous to \eqref{eq:mhatrelation}:
\begin{equation}
\label{eq:mtilderelation}
    \tilde{m}_{ij} \hs{4}J^j(x) = - J_i~.
\end{equation}
This differs from \eqref{eq:mhatrelation}
only by a missing prefactor 1/2.
With this, we can give the final form of the equation of motion which we will use:
\begin{equation}
    \codiff \left( \hat{M}_{ij}  F_3^j \right) 
    =  \epsilon_{ij} \hs{10} \left(F_3^j \wedge \tilde{F}_5\right)+ \left( J_i - \tilde{m}_{ij} C_2^j \right)  \pair{\delta}(y-y_{D3})~.
    \label{ueom}
\end{equation}

Now we turn on background fluxes $\bcgr{F_3}^i$, $\bcgr{\tilde{F}}_5$~for the 3- and 5-form fields strengths. To find the leading effects of these background fields we substitute $F_3^i\to \bcgr{F_3}^i + F_3^i$ and 
$\tilde{F}_5 \to \bcgr{\tilde{F}}_5 + \tilde{F}_5$ in \eqref{ueom} and keep only terms linear in the field  fluctuations $F_3^i$ and $\tilde{F}_5$
(see also \cite{Kim:1985ez,Ceresole:1999ht}): 
\begin{equation}
    \label{eq:final-C2-eom}
    \begin{aligned}
        (\hat{M}^{-1})^{ij} J_j ~\pair{\delta}(D3) = \left( \codiff \d C_2^i  + (\hat{M}^{-1})^{ij} \tilde{m}_{jk} C_2^k~ \pair{\delta}(D3)\right) \\&\hspace{-2.2cm}-(\hat{M}^{-1})^{ij}\epsilon_{jk}\hs{10}\left( \bcgr{F}_3^k \wedge \tilde{F}_5 + F_3^k \wedge \bcgr{\tilde{F}}_5\right)~.
    \end{aligned}
\end{equation}
An analogous procedure applied to the $C_4$ equation of motion \eqref{eq:eomC4} gives
\begin{equation}
    \label{eq:final-C4-eom}
    0=\codiff \tilde{F_5} - \epsilon_{ij} \hs{10}\left( \bcgr{F}_3^i \wedge F_3^j \right) 
    ~,
\end{equation}
where the background $\bcgr{\tilde{F}}_5$ has to satisfy
\begin{equation}\label{eq:F5-bcgr}
    \d \hs{10}\bcgr{\tilde{F}}_5=\frac{\epsilon_{ij}}{2} \bcgr{F}_3^i \wedge \bcgr{F}_3^j -\pair{\delta}_6(y-y_{D3})~.
\end{equation}
The delta distribution on the right hand side represents just one D3-brane but, of course, we have to imagine this being replaced by the full set of localised sources, including in particular O3-planes. Theoretically it is now straightforward to derive KM by simultaneously solving the equations of motion for $C_2^i$ \eqref{eq:final-C2-eom} and $C_4$ \eqref{eq:final-C4-eom} in the background of fluxes and localised sources.

To make progress towards an explicit result, we will now argue that, in the large volume limit, it is consistent to set $\bcgr{\tilde{F}}_5$ to zero. 
To do so, we first discuss separately the effects coming from two distinct regions: (A) the near-D3 regions with their strongly-peaked $\bcgr{\tilde{F}}_5$ profile and (B) the generic bulk region, where $\bcgr{\tilde{F}}_5$ represents a dilute flux background, suppressed at large volume.

As will be discussed in app.~\ref{app:coupling-renormalization}, the $\bcgr{\tilde{F}}_5$ effects from region (A) can, together with further effects related to $\tilde{m}_{ij}$, be absorbed in a renormalization of the brane action. Specifically, this will lead to an effective brane coupling to $C_2^i$, which we expect to deviate from the leading-order result at most by an ${\cal O}(1)$ factor. In fact, it will become clear that we may expect any such effects to be suppressed by $g_s$ if, in addition to being at large volume, we assume $g_s\ll 1$. This is sufficient for our purposes and we disregard $\bcgr{\tilde{F}}_5$ effects from region (A) for now.

In region (B), both $\bcgr{\tilde{F}}_5$ and $\bcgr{F}_3^i$ are dilute and we may consider an expansion in these backgrounds. At zeroth order in $\bcgr{F}_3^i$ and all orders in $\bcgr{\tilde{F}}_5$, no kinetic mixing arises by an SL$(2,\mathbb{R})$ argument similar to sec.~\ref{sc:GCwoF}. This is explained in app.~\ref{app:cancelling-diagrams}. At linear order in $\bcgr{F}_3^i$ and all orders in $\bcgr{\tilde{F}}_5$, no kinetic mixing arises. This will become clear when we discuss the corresponding diagrams below. Finally, going to quadratic order in $\bcgr{F}_3^i$, we will find a nonzero result already at zeroth order in $\bcgr{\tilde{F}}_5$. Any terms of quadratic order in $\bcgr{F}_3^i$ involving also $\bcgr{\tilde{F}}_5$ will then be subleading and we disregard them.

Thus, we now proceed setting 
$\bcgr{\tilde{F}}_5=0$. The only flux effect is then the mixing between $C_2^i$ and $C_4$, which arises from the terms $\sim \bcgr{F}_3^i$ in \eqref{eq:final-C2-eom} and \eqref{eq:final-C4-eom}.
To solve \eqref{eq:final-C2-eom} and \eqref{eq:final-C4-eom} together for $C_2^i$ and $C_4$, 
it proves useful to split the forms and derivatives in 4d and 6d parts:
\begin{gather}
    A_p = \sum\limits_{q=0}^p A_{(p-q,q)} ~,\\
   \d = \dfour + \dsix~,\\
   \codiff = \ddfour + \ddsix~.   
\end{gather}
Here e.g. $A_{(2,1)}$ corresponds to a 2-form in 4d and a 1-form in 6d.
Since the brane source $J_{i,(2,0)}$ is a pure 4d 2-form, we are only interested in the $(2,0)$ component of \eqref{eq:final-C2-eom}.
Assuming product form of the 10d metric we have\footnote{This can easily be deduced from $\codiff\d A_{\mu_1 \dots \mu_p}=-(p+1) \nabla^\alpha \nabla_{[\alpha}A_{\mu_1 \dots \mu_p]}$, see e.g.~App.~A in \cite{Hinterbichler:2013kwa}.}
\begin{equation}
    \codiff\d C_2^i  = (\ddfour \dfour + \ddsix \dsix ) C_2^i~.
\end{equation}
It follows that the operator $\codiff\d$ does not mix modes of e.g. $C_2^i$ with different 4d/6d form degrees.
Hence, given that we also assume $\bcgr{\tilde{F}}_5=0$, there is no mixing between such different $C_2^i$ modes in \eqref{eq:final-C2-eom}. The (2,0) component of \eqref{eq:final-C2-eom} thus reads
\begin{equation}
\begin{split}
        \label{eq:C20-eom}
        (\hat{M}^{-1})^{ij} J_{j,(2,0)} ~\pair{\delta}(D3) = \left(\delta_k^i~ \codiff \d   + (\hat{M}^{-1})^{ij} \tilde{m}_{jk} ~ \pair{\delta}(D3)\right) C_{(2,0)}^k \\&
        \hspace{-2.3cm}
        -(\hat{M}^{-1})^{ij}\epsilon_{jk}\hs{10}\left( \bcgr{F}_{(0,3)}^k \wedge \tilde{F}_{(2,3)} \right)~,
    \end{split}
\end{equation}
where we also used the fact that our 3-form flux is  internal: $\bcgr{F}^i_3=\bcgr{F}_{(0,3)}^i$.
All other modes of $C_2^i$ decouple and \eqref{eq:C20-eom} is the only equation we need to consider further. 

The only $C_4$ modes that couple in \eqref{eq:C20-eom} are $C_{(2,2)}$ and $C_{(1,3)}$ which can be seen by considering the definition of $\tilde{F}_{(2,3)}$,
\begin{equation}
    \tilde{F}_{(2,3)}=\d C_4\Big|_{(2,3)}=\dfour C_{(1,3)} + \dsix C_{(2,2)}~.
\end{equation}
Further simplifications arise if, as before, we neglect 4d derivatives of fluctuating fields relative to 6d derivatives. This implies
\begin{align}
    \tilde{F}_{(2,3)} &\approx\dsix C_{(2,2)}~,\\
    F^j_{(2,1)} &\approx \dsix C_{(2,0)}^j~,
\end{align}
which means that we can forget about $C_{(1,3)}$ since it decouples from $C^j_{(2,0)}$ in \eqref{eq:C20-eom}.
\footnote{
Note that, in general, only the combinations $\dfour C_{(1,3)}+\dsix C_{(2,2)}$ and $\dfour C^j_{(1,1)}+\dsix C_{(2,0)}^j$ are gauge invariant. However, fixing the gauge by \eqref{eq:C2-gauge} and \eqref{eq:C4-gauge} we can omit both $\dfour C^j_{(1,1)}$ and $\dfour C_{(1,3)}$ since $\dsix C_{(2,0)}^j$ and $\dsix C_{(2,2)}$ are invariant under residual gauge transformations left after the gauge choice \eqref{eq:C2-gauge}, \eqref{eq:C4-gauge}. Technically this becomes apparent after performing a Hodge decomposition of the form fields \cite{Hinterbichler:2013kwa}.
}
The equations of motion for the remaining $C_{(2,2)}$ mode follows from \eqref{eq:final-C4-eom} and reads
\begin{gather}
    \label{eq:C22-eom}
    0=\codiff \d C_{(2,2)} - \epsilon_{ij} \hs{10} \left(\bcgr{F}_{(0,3)}^i \wedge F^j_{(2,1)}\right) ~.
\end{gather}

We thus find a closed set of two equations of motion, \eqref{eq:C20-eom} and \eqref{eq:C22-eom}, which we re-write in matrix notation as
\begin{equation}\label{eq:final-eom-to-invert}
    \begin{pmatrix}
        -\hs{4}J^i_{(2,0)} \pair{\delta}(D3)\\0
    \end{pmatrix}
    =
    \hat{D}^i_{~k}
    \begin{pmatrix}
        C^k_{(2,0)}\\
        C_{(2,2)}
    \end{pmatrix}
    ~,
\end{equation}
where we defined 
\begin{gather}
    \hat{D}^i_{~k} = \begin{pmatrix}
        \delta_k^i~ \Delta_6   + (\hat{M}^{-1})^{ij} \tilde{m}_{jk} ~ \pair{\delta}(D3) & (\hat{M}^{-1})^{ij}P_j[\hspace{2pt}\cdot\hspace{2pt}]\\
        P_k[\hspace{2pt}\cdot\hspace{2pt}]&\Delta_6
    \end{pmatrix}
    ~,\\
    P_j[\hspace{2pt}\cdot\hspace{2pt}]=  \epsilon_{jk} \hs{10}\left( \bcgr{F}_{(0,3)}^k \wedge \dsix [\hspace{2pt}\cdot\hspace{2pt}] \right)~.
\end{gather}
Here we also replaced $\codiff\d \to \Delta_6$, which is justified if we impose the gauges \cite{Kim:1985ez,Ceresole:1999ht} 
\begin{gather}
    \label{eq:C2-gauge}
    \ddfour C_2^i = 0 = \ddsix C_2^i~,\\
    \label{eq:C4-gauge}
    \ddfour C_4 = 0 = \ddsix C_4~,
\end{gather}
and if we neglect 4d derivatives.

\subsection{Leading order result}\label{sc:LO-result}
\begin{figure}[h]
    \centering
    \subfloat[]{\includegraphics[height=3cm,valign=c]{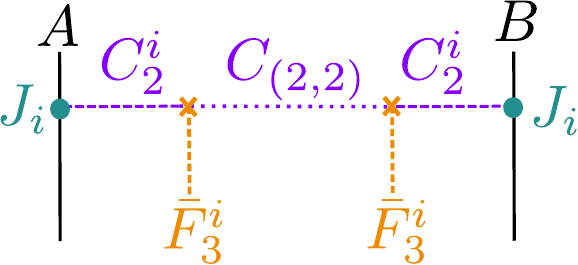}
    }
    \hspace{13pt}
    \subfloat[]{\includegraphics[height=3cm,valign=c]{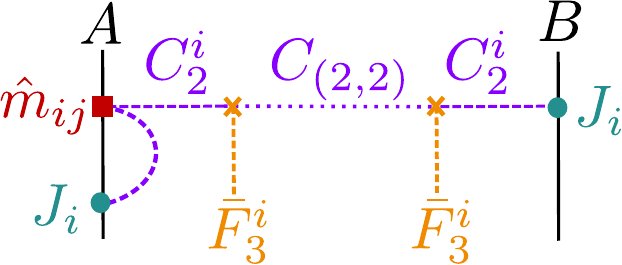}
    }
    \caption{Including the background for $\bcgr{F}^i_{3}$ introduces a new coupling term in the bulk. Thus, there are now more diagrams potentially contributing to KM. Diagram~(a) corresponds to the leading-order effect and (b) shows an example of a further diagram, involving $\hat{m}$ and contributing to KM at higher order in $g_s$ (cf.~app.~\ref{app:coupling-renormalization}).}
    \label{fig:flux-diagrams}
\end{figure}

We now want to solve \eqref{eq:final-eom-to-invert} in order to integrate out $(C^k_{(2,0)},~C_{(2,2)})$ to obtain a leading order result for KM.\footnote{
Note that the zero-mode of $C_{(2,2)}$ is \textit{not} projected out by orientifolding. So one may be concerned that $C_{(2,2)}$ can not be integrated out. However, in the key equations  \eqref{eq:final-eom-to-invert} only 6d derivatives of this field appear. Hence, the zero mode decouples and $C_{(2,2)}$ can be integrated out together with $C^k_{(2,0)}$.
}
Again we need to invert $\hat{D}^i_{~k}$ which we do by expanding in $\tilde{m}$ and $P_j$. For this purpose we decompose $\hat{D}^i_{~k}$ according to
\begin{gather}\label{eq:D-operator}
    \hat{D}^i_{~k} = (\hat{D}_{(0)})^i_{~k} + \delta\hat{D}^i_{~k}~,\quad
    (\hat{D}_{(0)})^i_{~k}=\begin{pmatrix}
        \delta_k^i~ \Delta_6    & 0\\
        0&\Delta_6
    \end{pmatrix}
    ~,\\[7pt]
    \delta\hat{D}^i_{~k}= 
    \begin{pmatrix}
         (\hat{M}^{-1})^{ij} \tilde{m}_{jk} ~ \pair{\delta}(D3) & (\hat{M}^{-1})^{ij}P_j[\hspace{2pt}\cdot\hspace{2pt}]\\
        P_k[\hspace{2pt}\cdot\hspace{2pt}]&0
    \end{pmatrix}~.
\label{dddef}
\end{gather}
Considering only $(\hat{D}_{(0)})^i_{~k}$, the analysis is equivalent to sect.~\ref{sc:LOC}. Including $\delta\hat{D}^i_{~k}$ but keeping only the term $\sim \,\tilde{m}$ 
in its definition, \eqref{dddef}, is equivalent to sect.~\ref{sc:GCwoF}. Both results were identically zero. Non-zero contributions arise once we include the $P_k$ terms from \eqref{dddef}. These are proportional to $\bcgr{F}^i_{(0,3)}$ and, thinking in terms of diagrams, we may associate them with 3-vertices involving the flux background and two propagating fields $C_2^i$, $C_{(2,2)}$.
This is illustrated in fig.~\ref{fig:flux-diagrams}.\footnote{
Strictly speaking, we do not have the right to base our analysis on Feynman diagrams since we do not have an action for fluctuations around a background with non-zero $\bcgr{F}_3$. However, as long as we stick to tree-level diagrams, we can view these as simple mnemonic rules for organizing the perturbative solution of the classical EOMs for $C_2^i$ and $C_4$. Thus, at our level of precision, we may use EOM and Feynman diagram language interchangeably.
}

\noindent
Inverting $\hat{D}$ as a power series in $\delta\hat{D}^k_{~l}$ we find
\begin{equation}
    (\hat{D}^{-1})^i_{~j} = (\hat{D}^{-1}_{(0)})^i_{~j} + (\hat{D}^{-1}_{(0)})^i_{~k} ~\delta\hat{D}^k_{~l}~ (\hat{D}^{-1}_{(0)})^l_{~j} +\cdots~,
\end{equation}
where the first-order term $\hat{D}^{-1}_{(1)}$ explicitly reads
\begin{equation}
    \begin{aligned}
    (\hat{D}^{-1}_{(1)})^i_{~j} = 
    &(\hat{D}^{-1}_{(0)})^i_{~k} ~\delta\hat{D}^k_{~l}~ (\hat{D}^{-1}_{(0)})^l_{~j} 
    \\= &
    \begin{pmatrix}
        \Delta_6^{-1}(\hat{M}^{-1})^{ik}\tilde{m}_{kj} [\pair{\delta}(A)+\pair{\delta}(B)] \Delta_6^{-1} & \Delta_6^{-1} (\hat{M}^{-1})^{ik}P_k \circ \Delta_6^{-1} \\
        \Delta_6^{-1} P_j \circ \Delta_6^{-1} & 0
    \end{pmatrix}~.
    \end{aligned}
\end{equation}
Integrating out the vector $(C^k_{(2,0)},~C_{(2,2)})^{\text{T}}$ at this order contributes to the KM action as
\begin{equation}\label{eq:KM-contribution-example}
    S\supset \frac{\bt_3}{2} \int\limits_{\extmfd}  \Big(J_{i,(2,0)}^{\fieldlable{(A)}}  ,~0 \Big)\whs{4} \left[ (\hat{D}^{-1}_{(1)})^i_{~j} \begin{pmatrix}
        -\hs{4}\pair{J}_{(2,0)}^{\fieldlable{(B)}\,j} \\0
    \end{pmatrix} \right] + (A \leftrightarrow B)~.
\end{equation}
Due to the 0 in the lower component of the source vector, only the upper-left entry of $\hat{D}^{-1}_{(1)}$ is relevant. We know, however, that this contribution vanishes by the arguments of sect.~\ref{sc:GCwoF}. Hence we have to go to the second order term $\hat{D}^{-1}_{(2)}$, where we again focus only on the top-left entry:
\begin{equation}\label{eq:snd-order-contribution}
    \begin{aligned}
    (\hat{D}^{-1}_{(2)})^i_{~j}\Big|_{\fieldlable{top-left}} &= 
    (\hat{D}^{-1}_{(0)})^i_{~k} ~\delta\hat{D}^k_{~l}~ (\hat{D}^{-1}_{(0)})^l_{~r} ~\delta\hat{D}^r_{~s}~ (\hat{D}^{-1}_{(0)})^s_{~j} \Big|_{\fieldlable{top-left}}
    \\ &
        =\Delta_6^{-1}(\hat{M}^{-1})^{ik}\tilde{m}_{kl} [\pair{\delta}(A)+\pair{\delta}(B)] \Delta_6^{-1}(\hat{M}^{-1})^{lr}\tilde{m}_{rj} [\pair{\delta}(A)+\pair{\delta}(B)] \Delta_6^{-1} \\
        &\qquad+ \Delta^{-1}_6 (\hat{M}^{-1})^{ik} P_k \Delta^{-1}_6 P_j \Delta^{-1}_6~.
    \end{aligned}
\end{equation}
Once again, the KM contribution from the first term in \eqref{eq:snd-order-contribution} cancels by the analysis of sect.~\ref{sc:GCwoF}. However, the second term in \eqref{eq:snd-order-contribution} provides a non-zero contribution. This turns out to be our leading-order result.
Similarly to \eqref{eq:KM-contribution-example}, we may write the corresponding action term as
\begin{equation}\label{eq:intermediate-KM-result}
    S\supset  \frac12\int\limits_{\extmfd} \left( J^{\fieldlable{(A)}}_{i,(2,0)} \whs{4} J^{\fieldlable{(B)}}_{j,(2,0)} ~K^{ji}(y_A,y_B)+J^{\fieldlable{(B)}}_{i,(2,0)} \whs{4} J^{\fieldlable{(A)}}_{j,(2,0)} ~K^{ji}(y_B,y_A) \right)~,
\end{equation}
where $K^{ji}$ is given by\footnote{We have integrated by parts to make the symmetry of $K^{ji}(y_A,y_B)$ more apparent.}
\begin{equation}\label{eq:KM-prefactor}
    \begin{aligned}
    K^{ji}(y_A,y_B)= 2\pi\int\limits_{y',y''}\Bigg( \frac{1}{4!} G_6(y_A,y') ~\bcgr{F}^{j}(y')_{[abc}& ~\partial_{d]}^{(y')}\partial^d_{(y'')}\left[G_6(y',y'')\right] \\
    &\times \bcgr{F}^i(y'')^{abc} ~G_6(y'',y_B)\Bigg) ~.
    \end{aligned}
\end{equation}
Here we introduced the scalar Green's functions $G_6$, to be distinguished from the general Green's function $\Delta^{-1}_6$ which acts on forms and is itself form-valued,  cf.~\cite{Thirring:1979bh, LechnerEdyn}. This feature is particularly relevant in connection with $C_{(2,2)}$.  The symbolic manipulations above are not affected by this technicality.
The labels $a,b,c,d$ refer to 6d indices and we used the identity \mbox{$2\kappa_{10}^2 \bt_3^2=2\pi$}.
Crucially, the argument from secs.~\ref{sc:LOC}
and \ref{sc:GCwoF} for the vanishing of KM does not apply since the tensor $K^{ij}$ used to contract $J_{i,(2,0)}^{(A)}$ and $J_{j,(2,0)}^{(B)}$ is, in contrast to $\epsilon^{ij}$, symmetric rather than antisymmetric:
\begin{equation}
    K^{ji}(y_A,y_B) = K^{ij}(y_B,y_A)~.
\end{equation}
We can then simplify \eqref{eq:intermediate-KM-result} to obtain the final result
\begin{equation}
    \label{eq:final-KM}
    S\supset \int\limits_{\extmfd} J^{\fieldlable{(A)}}_{i,(2,0)} \whs{4} J^{\fieldlable{(B)}}_{j,(2,0)} ~K^{ji}(y_A,y_B)~.
\end{equation}

\subsection{Open issues with the supergravity embedding}\label{sc:issues-with-sugra}

From \eqref{eq:KM-prefactor}, \eqref{eq:final-KM} we can already infer that our result will depend on the distance between the two D3-branes, which may be problematic as will become clear momentarily.
More specifically, as will be worked out in detail in the next section, $K^{ji}$ scales as $1/{\cal V}^{4/3}$ with ${\cal V}$ the Calabi-Yau volume in 10d Planck units. The difficulties arise because our 4d EFT is a genuine ${\cal N}=1$ supergravity theory.
Further, it has already been stated that KM arises as a loop correction to the gauge kinetic function $f_\fieldlable{AB}$ \cite{Benakli:2009mk,Goodsell:2009xc,Bullimore:2010aj}.
In supergravity, the gauge kinetic function is holomorphic. Thus, the volume dependence noted above implies that $f_\fieldlable{AB}$ is a holomorphic function of the Kahler moduli. However, the shift symmetry of the Kahler moduli excludes any holomorphic Kahler moduli depence of $f_\fieldlable{AB}$ which is not linear or exponential \cite{Benakli:2009mk,Goodsell:2009xc,Bullimore:2010aj}. This clashes with \eqref{eq:final-KM} and implies the question how our result can be understood from a 4d supergravity perspective.

In fact, our computation in the present paper was not loop-based but relied on the equivalent approach of integrating out 10d $p$-form fields. In 4d language, this corresponds to integrating out 
an infinite tower of heavy KK modes.
Equivalently, one may say that we are integrating out massive strings stretched between the branes.
Such a procedure of integrating out massive fields in supergravity is potentially problematic, as has been pointed out in \cite{Choi:2004sx,deAlwis:2005tf,deAlwis:2006ifb,deAlwis:2005tg,Brizi:2009nn}.
It is in particular plausible that it induces
higher derivative operators in the 4d effective theory, similar to those  discussed in \cite{Cecotti:1986jy,Cecotti:1986pk,Pickering:1996he,Khoury:2010gb,Baumann:2011nm,Koehn:2012ar,Cicoli:2013swa,Kuzenko:2014ypa,Ciupke:2015msa,Ciupke:2015ora,Bielleman:2016grv,Cicoli:2016chb,Ciupke:2016agp,Weissenbacher:2016gey,Grimm:2017pid,Grimm:2017okk,Cicoli:2023njy}.
Additionally, to obtain our non-zero result it is crucial to include 3-form fluxes to break SL$(2,\mathbb{R})$. Including such fluxes generically breaks
SUSY spontaneously. As a result, higher-derivative corrections to 4d supergravity,\footnote{
By this we mean both terms of the type $\int d^4\theta\, W^\alpha W_\alpha f(\Phi,\overline{\Phi})$, which induce higher-derivatives in the on-shell action, as well as terms involving higher SUSY derivatives.
} 
such as those in eqs. (3.19) -- (3.21) of \cite{Goodsell:2009xc} or (3.23) of \cite{Bielleman:2016grv}, can affect the gauge-kinetic function, inducing a shift-symmetric and non-holomorphic Kahler moduli dependence.
However, this logic also implies that our KM result \eqref{eq:final-KM} must vanish if only SUSY-preserving fluxes are present. While this might well be the case, it is unfortunately not obvious from our result. More work is necessary to clarify this point.

\section{Implications for phenomenology}\label{sc:pheno}
Before estimating the magnitude of KM on the basis of \eqref{eq:final-KM}, we have to mention a caveat: Strictly speaking, the D3-brane model we investigated is not phenomenologically interesting because it involves no light states charged under the two gauge groups. Hence, one can perform a field redefinition such that any KM visible at low energies disappears. However, even in this toy model the KM we calculated is in principle a well-defined physical observable. Indeed, the model contains heavy charged states in the form of strings stretched between our two D3-branes and the corresponding mirror D3$'$-branes. The latter must be present due to our use of an O3/O7 orientifold. Allowing for any number and type of such states, one fills out a complete, two-dimensional integer charge lattice. If one works in a gauge field basis defined by this integer lattice, the KM term is fixed in an unambiguous way. For example, one could obtain a static interaction potential between two heavy states one charged under U(1)$_{\fieldlable{(A)}}$ the other under U(1)$_{\fieldlable{(B)}}$. In this case the interaction potential would be proportional to the KM. In this sense we claim that KM in our model is physical since charged states, even though heavy, are present.
However, we note that this is slightly problematic from the effective field theory point of view since these heavy states are not clearly much lighter than the states we have integrated out to obtain the kinetic mixing. 
Clearly, a better model would contain light charged states, which could be realized by considering branes at singularities or intersecting branes. For the singularity case, a string loop calculation has been performed in a particular class of torus orbifold models \cite{Bullimore:2010aj}, but we would need an appropriate 10d supergravity analysis. We leave this to future work. A third way of including light states will be discussed at the end of this section since it will benefit from formulae we will derive momentarily.

Thus, let us continue with the analysis of our example of single D$3$-branes. We will set $2\pi \sqrt{\ap}=1$ from now on. It will be convenient to introduce a length-scale-type variable $R$ associated with the volume $\cal V$, measured in 10d Einstein frame. We use the torus-motivated definition
\begin{equation}\label{eq:volume-relation}
    \V = (2\pi R)^6~,
\end{equation}
but we will think of $R$ more generally as of a typical length scale of our Calabi-Yau.
For a parametric estimate of KM from \eqref{eq:final-KM}, we need to characterize the magnitude of fluxes and Green's functions.  The 3-form flux $\bcgr{F}_{(0,3)}^i$ satisfies the standard quantization condition when integrated over a 3-cycle $\Sigma_3$ \cite{Giddings:2001yu}:
\begin{equation}\label{eq:fluxQuant}
    \int\limits_{\Sigma_3} \bcgr{F}_{(0,3)}^i =n^i\in \mathbb{Z}~.
\end{equation}
Using $\vol{\Sigma_3}\simeq\V^{1/2}$ this implies
\begin{equation}\label{eq:flux}
    \bcgr{F}_{mno}^i\sim  n^i ~\frac{1}{\V^{1/2}} ~.
\end{equation}
In the regime $y\ll \V^{1/6}$, the Green's function on the Calabi-Yau $\Delta_6^{-1}(y)$ can be estimated on the basis of its flat-6d counterpart (see \cite{Junghans:2023lpo} for further discussion)
\begin{equation}\label{eq:greensfunctionScaling}
    \Delta_6^{-1}(y) \simeq -\frac{1}{4\pi^3~y^4}~,\quad
    \partial_{(y)}\partial_{(y)}\Delta_6^{-1}(y) \simeq -\frac{5}{\pi^3 y^6}~.
\end{equation}

Using \eqref{eq:flux} and \eqref{eq:greensfunctionScaling} we may now estimate $K^{ji}$ from its definition in \eqref{eq:KM-prefactor}. In doing so, we will {\it not} implement the imaginary self-duality (ISD) condition \cite{Giddings:2001yu}
\begin{equation}
    i(F_3 -\tau H_3) =  \hs{6} (F_3 -\tau H_3) \qquad\Leftrightarrow\qquad F_3 = -\gs \hs{6} H_3+C_0H_3\,.
\end{equation}
As a result, we will also not be able to keep track of the $\gs$ dependence introduced by the ISD condition. This would require keeping track of the $g_s$ dependence of the relative size of 3-cycles, which appears due to relations like (for $C_0 = 0$)
\begin{equation}
    \int\limits_{\Sigma_3} F_3 = \int\limits_{\Sigma_3} (-\gs) \hs{6} H_3 = n\in \mathbb{Z}\,.
\end{equation}
Controlling the metric at this level of precision goes beyond our goals in this paper.

For our following simple estimates, we set $n^1=n^2=1$ and disregard the non-trivial profiles of the different fluxes on the Calabi-Yau.
The behaviour of the Green's functions now comes into play when we try to estimate the integrals defining $K^{ji}$ in \eqref{eq:KM-prefactor}. Crucially, one finds that these integrals are IR dominated, i.e.~the integrand does not diverge at $y'\rightarrow0$ and/or $y''\rightarrow0$.
We may then estimate $K^{ji}$ by inserting the maximal distance $y=\pi R= \V^{1/6}/2$ for $y'$ and $y''$ into the Green's functions \eqref{eq:greensfunctionScaling}, though this flat-space formula is at this point of course at best correct at the ${\cal O}(1)$ level. Replacing the integration by multiplication with the volume of the integration domain, one finds\footnote{
A corresponding diagrammatic calculation appears in
at the beginning of sect.~\ref{app:vol-sup-diagrams}}.
\begin{equation}\label{eq:Kij-scaling}
    K^{ji} \sim -\frac{2\pi}{4!} \frac{2^{10}}{\pi^9} \frac{5}{\V^{4/3}}~.
\end{equation}
We may be slightly more precise by ascribing different (though always ${\cal O}(1)$) numbers to the RR and NS fluxes:
$\bcgr{F}^{(1)}\sim f$ and $\bcgr{F}^{(2)}\sim h$.
This gives
\begin{equation}
    K^{ji} \sim -\frac{2\pi}{4!} \frac{2^{10}}{\pi^9} \frac{5}{\V^{4/3}} \,
    \begin{pmatrix}
        f^2 & f h\\
        f h & h^2
    \end{pmatrix}^{ji}\,.
\end{equation}
Note that $f$ transforms as a pseudoscalar, which follows from the CP properties of $F_3$.

We now turn to the product $J_i \whs{} J_j$ in \eqref{eq:final-KM}. By the definition of $J_i$ in \eqref{eq:F-sources} this will introduce $\gs$ and $C_0$ factors. A final subtlety arises because KM is defined with  canonically normalised gauge field strengths $\mathbb{F}_2$, cf. \eqref{eq:kinetic-mixing-def}.
By contrast, the field strength in our stringy analysis, $F_2$ from \eqref{eq:DBI-action}, is normalised by the coupling to the open string. The relation between the two reads
\begin{equation}
    \mathbb{F}_2=\gs^{-1/2}\sqrt{\bt_3} ~F_2~.
\end{equation}
Considering all the above details, the result for the parametric scaling of \eqref{eq:final-KM} is
\begin{equation}\label{eq:KM-estimate}
    \begin{aligned}
        S\,\,\supset\,\,\sim\int\limits_{\extmfd} \frac{2^{11}~5}{4!\pi^{9}~\V^{4/3}}\Bigg\{~ & \mathbb{F}_2^\fieldlable{(A)} \whs{4} \mathbb{F}_2^\fieldlable{(B)} &&\Big[ f^2 \gs + 2 fh ~\gs C_0 + h^2~(\gs^{-1} - \gs C_0^{~2})\Big]\\
        &\hspace{-5pt}+\mathbb{F}_2^\fieldlable{(A)} \wedge \mathbb{F}_2^\fieldlable{(B)} &&\Big[2 fh-2h^2~ C_0\Big]\Bigg\}~.
    \end{aligned}
\end{equation}
Comparing to \eqref{eq:kinetic-mixing-def}, we find the following estimates for the kinetic mixing parameter $\kmp_{\fieldlable{AB}}$ and the magnetic mixing (MM) parameter $\tilde{\kmp}_{\fieldlable{AB}}$:
\begin{gather}\label{eq:KM-scaling}
    \kmp_{\fieldlable{AB}}\sim -\frac{2^{10}}{4!\pi^{9}}~ \frac{5\gs^{-1}}{\V^{4/3}}~,\\
    \label{eq:MM-scaling}
    \tilde{\kmp}_{\fieldlable{AB}} \sim -\frac{2^{11}}{4!\pi^{9}} \frac{5(f-  C_0)}{\V^{4/3}}\,.
\end{gather}
Here we also included the factor $1/2$ from the definition \eqref{eq:kinetic-mixing-def} and we set $h=1$.
We left $f$ explicit such that the CP properties become apparent. The main suppression of KM and MM will be due to the volume factor. We caution the reader that, while we tried to keep track of factors of $\pi$ in our estimates, the prefactor is nevertheless uncertain at the level of one or two orders of magnitude. This can be seen e.g. by using a $6$-torus Green's function \cite{CourantHilbert,Shandera:2003gx,Andriot:2019hay,Junghans:2023lpo} instead of a flat approximation \eqref{eq:greensfunctionScaling}, which introduces several $\pi$ factors due to the six internal dimensions. We indicate this uncertainty in fig.~\ref{fig:KM-Parameter-Plots} with the color shaded band around the bounds.

In order to obtain a parametric estimate of the smallest possible values for KM and MM, we now consider the implementation of our model in the large volume scenario (LVS) \cite{Balasubramanian:2005zx,Conlon:2005ki}.
In the LVS, we can constrain the maximal size of the internal geometry $\V$ that can be stabilized consistently.
The main constraint comes from the volume modulus, since the volume modulus is the lightest modulus and couples like gravity to all matter fields after Weyl rescaling to 4d Einstein frame.
In the LVS, the mass $m_\mathcal{V}$ in units of the 4d Planck mass $\Mpl$ is given by \cite{Conlon:2005ki}
\begin{equation}
\label{eq:MV-Vol-scaling}
    m_{\mathcal{V}}\sim \frac{ W_0}{\sqrt{4\pi} \gs^{1/4} \V^{3/2}}\Mpl
    ~.
\end{equation}
Thus we can rewrite \eqref{eq:KM-scaling} and \eqref{eq:MM-scaling} in terms of $m_\V/\Mpl$:
\begin{gather}
    |\kmp_{\fieldlable{AB}}|\sim~5\frac{2^{10}}{4!\pi^{9}} (4\pi)^{4/8} \left(\frac{m_\V}{\Mpl} \right)^{8/9}\gs^{-7/9}~ W_0^{-8/9}~,\\
    |\tilde{\kmp}_{\fieldlable{AB}}| \sim~5\frac{2^{11}}{4!\pi^{9}} (4\pi)^{4/8} \left(\frac{m_\V}{\Mpl} \right)^{8/9}\gs^{2/9}~ W_0^{-8/9} (f-C_0)~.
\end{gather}
A very conservative constraint for the volume modulus mass follows by demanding that fifth force limits~\cite{Kapner:2006si,Damour:2010rp} are respected, which implies $m_\V \gtrsim 10^{-30}\Mpl$. We can use this constraint on $m_\V$
to give a lower bound on $\kmp_{\fieldlable{AB}}$ and $\tilde{\kmp}_{\fieldlable{AB}}$
\begin{gather}
    \label{eq:bound1}
    |\kmp_{\fieldlable{AB}} | \gtrsim~ 2.8 \times 10^{-28} \left(\frac{m_\V}{\Mpl} \frac{1}{10^{-30}}\right)^{8/9}\left(\frac{\gs}{0.1}\right)^{-7/9} \left(\frac{W_0}{1}\right)^{-8/9}~,\\
    \label{eq:bound2}
    |\tilde{\kmp}_{\fieldlable{AB}}| \gtrsim~ 5.7 \times 10^{-29} \left(\frac{m_\V}{\Mpl} \frac{1}{10^{-30}}\right)^{8/9} \left(\frac{\gs}{0.1}\right)^{2/9} \left(\frac{W_0}{1}\right)^{-8/9} \frac{f-C_0}{1}~.
\end{gather}
Note that $W_0$ is constrained in terms of the D3-brane tadpole $Q_3\sim{\cal O}(100)$ \cite{Denef:2004ze}:
\begin{equation}
    W_0 \lesssim \sqrt{\frac{|Q_3|}{  \gs}}~.
\end{equation}
In the LVS, we further have the relation $\gs \sim  1/\ln\V$, which excludes large or extremely small values of $\gs$,
given that we insist on a theoretically and phenomenologically reasonable value for $\cal V$. We can hence not use the in principle tunable parameters $\gs$ and $W_0$ to reduce the lower bounds \eqref{eq:bound1} and \eqref{eq:bound2} significantly. As can be seen in fig.~\ref{fig:KM-Parameter-Plots}, the bound \eqref{eq:bound1} is well below the experimentally excluded region.

Tighter constraints on $m_\V$ may be derived from cosmology.
To avoid changing the element abundances by energy injection during BBN the modulus should decay well before this time~\cite{Coughlan:1983ci,Banks:1993en,deCarlos:1993wie}.
In addition, the modulus decays should not deposit a significant amount of energy into its own, ultra-light axion in order to avoid an excess of dark radiation~\cite{Cicoli:2012aq,Higaki:2012ar,Conlon:2013isa,Hebecker:2014gka,Angus:2014bia,Allahverdi:2014ppa,Cicoli:2015bpq,Cicoli:2018cgu,Acharya:2019pas,Angus:2021jpr,Jeong:2021yol,Frey:2021jyo,Cicoli:2021tzt,Cicoli:2022fzy}. 
Both problems can be circumvented if the volume modulus can decay efficiently into SM Higges~\cite{Cicoli:2022fzy}. This, in turn, requires the mass $m_\V $ to be large enough, $m_\V\gtrsim 2 m_H $, where $m_H$ refers to the Higgs mass. Using this bound yields
\begin{gather}
    \label{eq:bound1.2}
    |\kmp_{\fieldlable{AB}}| \gtrsim~ 8.1 \times 10^{-16} \left(\frac{m_\V}{\Mpl} \frac{\Mpl}{2m_H}\right)^{8/9}\left(\frac{\gs}{0.1}\right)^{-7/9} \left(\frac{W_0}{1}\right)^{-8/9}~,\\
    \label{eq:bound2.2}
    |\tilde{\kmp}_{\fieldlable{AB}}| \gtrsim~ 1.6 \times 10^{-16} \left(\frac{m_\V}{\Mpl} \frac{\Mpl}{2m_H}\right)^{8/9} \left(\frac{\gs}{0.1}\right)^{2/9} \left(\frac{W_0}{1}\right)^{-8/9} \left(\frac{f-C_0}{1}\right)~.
\end{gather}
Notably, values of the order of the bound \eqref{eq:bound1.2} are now being probed by experiments and observations, see fig.~\ref{fig:KM-Parameter-Plots}.

We emphasize again that, in generic flux compactifications where the complex structure moduli are stabilized along the line of \cite{Giddings:2001yu}, both $\bcgr{F}_{(0,3)}$ and $\bcgr{H}_{(0,3)}$ are turned on. One then expects both kinetic and magnetic mixing to be present. At least at the level of our simple single-D3-brane toy model, both kinetic and magnetic mixing vanish exactly if no 3-form fluxes are turned on.

From eqs.~\eqref{eq:bound1} and \eqref{eq:bound2} we infer that quite small values for KM are achievable without any tuning of the relevant gauge couplings. In particular, the generic estimate that ${\cal O}(1)$ gauge couplings imply $\kmp\sim {\cal O}(1)$ can be easily avoided.\footnote{
We 
note an apparent tension between our values and bounds on KM inferred from positivity constraints on gravitational scattering amplitudes argued for in~\cite{Aoki:2023khq} in an explicit Standard Model context. It would be interesting to match the two settings in detail and try to understand and resolve any possible discrepancy.}

We now want to return to the third possibility of including charged light states mentioned at the beginning of this section: We may replace our two D3-branes by two stacks of D3-branes, with the branes in each of them separated by a small distance $d$. Light charged states now arise from the strings stretched between the branes in each stack.
For concreteness, let each stack consist of two D3-branes. The light states are charged under the ``relative'' U$(1)^\fieldlable{(r)}$ which originates from breaking the brane stack gauge group according to U$(1)^\fieldlable{(o)}\times \text{SU}(2)\rightarrow \text{U}(1)^\fieldlable{(o)}\times \text{U}(1)^\fieldlable{(r)}$. The additional ``overall'' U$(1)^\fieldlable{(o)}$ has no light charged states. The two relative U$(1)$ gauge groups of the two stacks will then mix due to $C_2^i$-exchange, as analysed in detail in the bulk of this paper.

Due to the breaking of a non-abelian gauge group, we expect a further suppression factor to come into play. This is most easily seen from a field theory perspective. To make our point we focus on the even simpler case where only one of the two relevant gauge groups is non-abelian, e.g.~U(1)$_{\fieldlable{A}}$ and SU(2)$_{\fieldlable{B}}$. Before SU(2)$_{\fieldlable{B}}$ is broken, KM is clearly impossible because of the non-abelian structure. The leading operator governing KM must involve the SU(2)$_{\fieldlable{B}}$-breaking VEV and takes the form\cite{Goldberg:1986nk,Arkani-Hamed:2008kxc,Brummer:2009oul,Gherghetta:2019coi}
\begin{equation}
\mathcal{L} \supset  \frac{\kmp_\fieldlable{AB,0}}{\Lambda}  F_{\fieldlable{A}} \tr( \Phi_{\fieldlable{B}} F_{\fieldlable{B}})\,.
\end{equation}
Here $\kmp_\fieldlable{AB,0}$ is a parameter specifying any a-priori suppression of the interaction between the two gauge groups, as it arises in our context because of sequestering within the large CY volume. Moreover, $\Phi_\fieldlable{B}$ is an adjoint scalar and $\Lambda$ the UV cutoff scale. Thus, after breaking SU(2)$_\fieldlable{B}$ the KM mixing between U(1)$_{\fieldlable{A}}$ and the surviving U(1) from SU(2)$_\fieldlable{B}$ is governed by
\begin{gather}
\kmp_{\fieldlable{AB}} \sim \kmp_\fieldlable{AB,0}\frac{\vev{\Phi_\fieldlable{B}}}{\Lambda}~.
\end{gather}
Our key point here is the additional suppression by $\vev{\Phi_\fieldlable{B}}/\Lambda$.

In an analogous stringy setup, with one single D3 brane and a U(2)-stack with adjoint breaking, we  find
\begin{equation}
    \kmp_{\fieldlable{AB}} \sim \kmp_\fieldlable{AB,0} ~\frac{\gs^{-1/4}}{\V^{1/6}}~ \frac{\vev{\Phi}}{\stsc}\,,
\end{equation}
with $\vev{\Phi}\sim d ~\stsc$ and $d$ the brane-separation in the U(2) stack in string units. The derivation is given in app.~\ref{app:toy-model}.

In the formula above, $\kmp_\fieldlable{AB,0}$ is the KM parameter as we derived it for two single D3 branes at a large distance. 
We have also extracted a factor implementing a suppression (for $d\ll 1)$ by the adjoint VEV, as expected on EFT grounds. Interestingly, our stringy realization displays a further suppression factor $g_s^{-1/4}/\V^{1/6}$, corresponding to one power of the inverse CY radius in string units. Intuitively, this can be explained by the fact that $C_2^i$ now couples to the relative U$(1)^\fieldlable{(r)}$ of the stack, i.e. U$(1)^\fieldlable{(1)}-\text{U}(1)^\fieldlable{(2)}$. Thus, one is dealing with a dipole coupling in comparison to the monopole-type coupling we were discussing before.
Even though these considerations contribute positively to our goal of small KM, we emphasize that this model can not realise chiral matter and hence can not be made fully realistic. 
A SM sector requires more involved constructions, e.g. branes at singularities or intersecting branes.

\section{Conclusions}\label{sc:conclusion}

Dark or hidden photons featuring a small but non-vanishing mixing with the ordinary photon can be probed at a level of mixing angles of sometimes better than $\sim 10^{-15}$. This makes them an ideal tool to probe sequestered or hidden sectors that are often present in string theory. 
Therefore, understanding how very small values of kinetic mixing can arise in relevant string setups is of significant phenomenological interest. In addition, it is of theoretical interest in the sense that it tests our ability to realize an extremely small coefficient for an operator which should naively be present at the ${\cal O}(1)$ level.

In this paper we have analysed kinetic mixing between the gauge groups of two D3 branes, as it occurs due to the propagation of $B_2$ and $C_2$ through the bulk of the relevant Calabi-Yau orientifold. Earlier investigations~\cite{Abel:2003ue,Abel:2008ai,Goodsell:2009xc} found a leading-order cancellation between the $B_2$ and the $C_2$ contribution both in a string theory and in a 10d supergravity calculation. However, an additional coupling arises if a non-zero value for $C_0$ is taken into account. Moreover, there exists a coupling between $B_2$ and $C_2$ localized on the D3-branes.
As one of our key results, we demonstrate that an exact cancellation persists after taking these two effects into account. For this it was essential to include a term in the D3-brane action which, while known in principle, is missing in standard textbooks, cf. app.~\ref{app:ExtraB2C2term}. Using the complete action for the D3-brane, we can further tie the generalised cancellation to the SL$(2,\mathbb{R})$ symmetry of type IIB supergravity, which in particular acts as a self duality group on the gauge theory living on the D3-brane.

We then extended our discussion by allowing for non-vanishing 3-form fluxes, which break SL$(2,\mathbb{R})$ spontaneously.
In this case non-vanishing kinetic mixing is present. Moreover, given that SL$(2,\mathbb{R})$ is the self duality group of the D3-brane gauge theory, it is not surprising that both kinetic and magnetic mixing arise. Any mixing obtained in this setting is small due to two effects: First, independently of any cancellation the $B_2$ and $C_2$ propagation leads to a suppression of the mixing by the Calabi-Yau volume. One may call this a sequestering effect. On top of that, the necessary presence of fluxes and the dilution of the latter when the Calabi-Yau grows large leads to a further volume suppression.  We provide explicit, SL$(2,\mathbb{R})$-covariant formulae for kinetic and magnetic mixing.

Explicitly, both kinetic and magnetic mixing are suppressed by ${\mathcal{V}}^{-4/3}$, where $\mathcal{V}$ is the Calabi-Yau volume. Specifically in the Large Volume scenario, the parameter $\V$ is linked to the volume modulus mass $m_{\mathcal{V}}\sim \V^{-3/2}M_{\text{Pl}}$, which is subject to experimental constraints.
Based on this, we derived lower bounds on kinetic mixing. While it is intriguing that the resulting values of the mixing parameter fall in a range interesting for future probes, we have to recall that our bounds are only indicative because our single-D3-brane model is not realistic.

At this point we recall an open issue raised by our findings. Our approach was based on integrating out form fields in the 10d theory. Nevertheless, the results should be in line with an effective description in terms of 4d supergravity. In this language holomorphicity combined with the shift symmetry of Kahler moduli seemingly forbid a volume dependence of the form we found. However, such a dependence may arise from higher-derivative corrections to the 4d supergravity action if supersymmetry is spontaneously broken. This connection between SUSY-breaking and kinetic mixing is not apparent from our result and deserves further study.

Clearly, a more detailed phenomenological discussion should be based on realistic models, containing both light charged states and chiral matter. This requires the extension of our analysis to $U(1)$ gauge groups on branes at singularities or on intersecting branes.
Further, we did not introduce a mass terms for the dark photon, which can potentially change the kinetic mixing. We leave these generalizations to future work.

\section*{Acknowledgments}
We thank Jonathan Heckman, Ralph Blumenhagen and Timo Weigand for valuable discussions and comments. We are particularly grateful to Stefan Theisen for an exchange concerning the D3-brane action and the presence of the extra term in the WZ-action as well as to Joseph Conlon and Mark Goodsell for discussions about the challenges of a 4d supergravity formulation.  The work of AH is supported by the Deutsche Forschungsgemeinschaft (DFG, German Research Foundation) under Germany’s Excellence Strategy EXC 2181/1 - 390900948 (the Heidelberg STRUCTURES Excellence Cluster).
JJ is grateful for the support from the European Union’s Horizon 2020 research and innovation programme under the Marie Sklodowska-Curie grant agreement No 860881-HIDDeN.
RK is  supported by the International Max Planck Research School for Precision Tests of Fundamental Symmetries (IMPRS-PTFS).

\appendix

\section[Precise formulation of the type IIB and D\boldmath$p$-brane action]{Precise formulation of the type IIB\\ and D\boldmath$p$-brane action}
\label{app:ExtraB2C2term}
The low energy limit of type-IIB string theory is $\mathcal{N}=2$ type-IIB supergravity. It is defined by a set of covariant equations of motions\cite{Schwarz:1983qr, Howe:1983sra}.
Since the theory contains a 4-from with self-dual field strength, these equations do not follow from a manifestly Lorentz-invariant action \cite{Marcus:1982yu}.
Famously, this issue can be avoided by imposing self-duality as a constraint after varying the action $S_{\text{IIB}}$, 
which we repeat here for better readability~\cite{Bergshoeff:1995as,Bergshoeff:1995sq},
\begin{gather}
\begin{aligned}
\label{eq-APP:IIBaction}
S_{\text{IIB}} = \frac{1}{2\kappa_\text{10}^2}\int\limits_{\M} \dx{10}{x} &\sqrt{-G_E}~ \left( R_E - \frac{\partial_M \Bar{\tau} \partial^M \tau}{2 (\Im ~\tau)^2} \right) \\
&+\frac{1}{2\kappa_\text{10}^2}\int\limits_{\M} \left( -\frac{\hat{M}_{ij}}{2} F^i_3 \whs{} F^j_3 -\frac{1}{4} \tilde{F}_5 \whs{} \tilde{F}_5 - \frac{\epsilon_{ij}}{4} C_4\wedge F^i_3 \wedge F^j_3 \right)~.
\end{aligned}
\end{gather}
We also restate the field strength
\begin{equation}
    \tilde{F}_5 = \d C_4 -\frac12 C_2 \wedge \d B_2 + \frac12 B_2 \wedge \d C_2~,
\end{equation}
referring the reader back to sect.~\ref{sc:SL2R-notations} for the other definitions.

An additional information, which is apparently often overlooked or implicitly understood, is that when writing down the SL$(2,\mathbb{R})$-invariant action \eqref{eq-APP:IIBaction} one has performed a field redefinition of the original stringy fields, associated to the massless modes of the quantized string\cite{Bergshoeff:1995as, Bergshoeff:1995sq, Ortin:2015hya}.
In the following, we will specify these stringy fields by using a hat: ``$\,\hat{~}\,$''.
The field redefinitions include the transformation of the string frame metric $\hat{G}$ to the Einstein frame metric $G_E$ as well as of the stringy 4-form gauge potential $\hat{C}_4$,\footnote{A detailed discussion can be found in \cite{Ortin:2015hya} or the appendix of \cite{Kiritsis:2019npv} for the bosons and in \cite{Gheerardyn:2001jj} for the fermions.}
\begin{equation}
    \label{eq:StringyC4}
    C_4=\hat{C}_4-\frac12 \hat{B}_2\wedge \hat{C}_2~.
\end{equation}
Since the other fields are not transformed and in particular $\hat{B}_2=B_2$ and $\hat{C}_2=C_2$, we drop the hat for those fields. We note that $\hat{C}_4$ is not SL$(2,\mathbb{R})$ invariant,
\begin{equation}
    \label{eq:Chat-SL-trafo}
    \hat{C}_4' = \hat{C}_4 + \frac{1}{2} \begin{pmatrix} C_2~,&\hspace{-5pt}B_2   \end{pmatrix} \wedge \begin{pmatrix} a c & c b\\ c b & b d   \end{pmatrix}\begin{pmatrix} C_2\\B_2   \end{pmatrix}~,
\end{equation}
which explains why $C_4$ is often used even when working in the string frame.

Particularly relevant for us is the
gauge-invariant field strength $\tilde{F}_5$ expressed in terms of the stringy field $\hat{C}_4$ \cite{Green:1996bh,Cederwall:1996ri,Bergshoeff:1997cf,Bergshoeff:2001pv}\,,
\begin{equation}
    \tilde{F}_5 = \d \hat{C}_4 - C_2 \wedge \d B_2 ~,
\end{equation}
which makes it clear that $\hat{C}_4$ is gauged only by $C_2$,
\begin{equation}
    \label{eq:Chat-Trafo}
    \delta \hat{C}_4 = \d \Lambda_3 +  \Lambda_1^{(1)} \wedge \d B_2~.
\end{equation}
This observation is important since we need a consistent description of the IIB bulk theory together with D$p$-branes.
The Einstein-frame D$p$-brane action reads \cite{Fradkin:1985qd,Abouelsaood:1986gd,Dai:1989ua,Leigh:1989jq,Witten:1995im,Polchinski:1995mt,Li:1995pq,Douglas:1995bn,Bershadsky:1995qy,Green:1996bh,Green:1996dd,Witten:1996hc,Mourad:1997uc,Cheung:1997az,Minasian:1997mm}
\begin{gather}
\label{eq:DBraneAction}
S_\text{D$p$-brane} = S_\fieldlable{DBI} + S_\fieldlable{WZ}~,\\[5pt]
\label{eq:fullDBI}
S_\fieldlable{DBI} = -\bt_p \int\limits_{\W_{p+1}} \dx{p+1}{\xi} \sqrt{-\det \left(G^E_{ab} - e^{-\phi/2} B_{ab}+ e^{-\phi/2} F_{ab}\right)}~,
\\[-5pt]
\label{eq:ChernSimonsTerms}
S_\fieldlable{WZ}=\pm \bt_p \int\limits_{\W_{p+1}} \exp\left(F_2 - B_2\right)\wedge \sum\limits_q \hat{C}_q  \wedge \sqrt{\frac{\hat{A}(4\pi^2 \ap R_T)}{\hat{A}(4\pi^2 \ap R_N)}}~,
\end{gather}
where we in particular follow the conventions of \cite{Green:1996bh}.
We see that $S_\fieldlable{WZ}$ is built using the stringy RR-fields $\hat{C}_q$. The corresponding gauge-invariant field strengths are\footnote{
Note 
that $\tilde{F}_3\neq F_3\equiv dC_2$.
} 
\begin{equation}\label{app-eq:gaugeinv-fieldstrength}
     \tilde{F}_{q+1} = \d \hat{C}_q - \d B_2 \wedge \hat{C}_{q-2}~,
\end{equation}
with the $\hat{C}_q$ gauge-transforming as
\begin{equation}
    \label{app-eq:gaugetrafo}
    \delta \hat{C}_q = \d \Lambda_{q-1} - \d B_2 \wedge \Lambda_{q-3}~.
\end{equation}

The brane action \eqref{eq:DBraneAction} is gauge-invariant under \eqref{eq:C2B2-gaugetrafo} and \eqref{eq:Chat-Trafo}.
The combination $F_2-B_2$ is gauge invariant due to \eqref{eq:F2-gaugetrafo}.
Furthermore, the action \eqref{eq:DBraneAction} is SL$(2,\mathbb{R})$ self-dual as defined by \eqref{eq:tau-Trafo}, \eqref{eq:C2B2SL2RTrafo}, \eqref{eq:F2-SL-trafo} and \eqref{eq:Chat-SL-trafo}.

Most importantly, this means that when using the standard type-IIB bulk action as in \eqref{eq-APP:IIBaction} together with a brane action, then $\hat{C}_4$ in the latter has to be replaced by $C_4$
using \eqref{eq:StringyC4}.
Thus, \textit{an extra term $+\frac{1}{2} B_2\wedge C_2$ appears for every $C_4$}. As explained, this is necessary for gauge invariance of any brane and for SL$(2,\mathbb{R})$ self-duality of the D3.

The explanations above have no claim to originality and follow from carefully reading the literature \cite{Ortin:2015hya,Kiritsis:2019npv,Douglas:1995bn,Morrison:1995yi,Tseytlin:1996it,Bergshoeff:1996cy,Kimura:1999jb,Bergshoeff:2006gs}.
We checked in detail the self-consistency of our notation.\footnote
{
A different convention one occasionally encounters uses the combination $F_2 +B_2$ instead of $F_2-B_2$. Such a sign flip for $B_2$ shows up in \eqref{app-eq:gaugeinv-fieldstrength}, \eqref{app-eq:gaugetrafo} and the SL$(2,\mathbb{R})$ transformations \eqref{eq:C2B2SL2RTrafo}. Through the expressions for $\tilde{F}_3$ and $\tilde{F}_5$ it then also affects the bulk action.}

\section{Higher-order diagrammatics}\label{app:issues}

As explained in sect.~\ref{sc:feom}, our result for KM in \eqref{eq:final-KM} holds only at leading order. 
The presence of $\hat{m}$ and $\bcgr{\tilde{F}}_5$ induces a whole set of additional diagrams which must be taken into account. We can organize them in three different classes: Diagrams which cancel among themselves, diagrams which induce a renormalisation of brane couplings, and diagrams which give volume-suppressed contributions and can hence be neglected in a controlled approximation.

\subsection{Cancelling diagrams}\label{app:cancelling-diagrams}
\begin{figure}[h]
    \centering
    \subfloat[]{\includegraphics[height=2.5cm,valign=c]{Pictures/BraneDiagrams-m-exchange1.pdf}
    \label{fig:example-cancel-diagram-1}
    }
    \hspace{30pt}
    \subfloat[]{\includegraphics[height=2.5cm,valign=c]{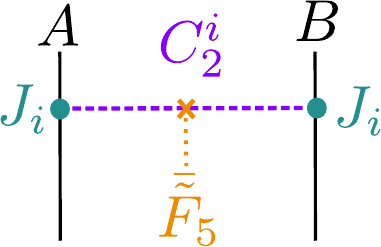}
    \label{fig:example-cancel-diagram-2}
    }
    \\
    \subfloat[]{\includegraphics[height=2.5cm,valign=c]{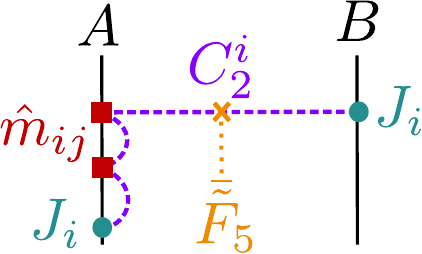}
    \label{fig:example-cancel-diagram-3}
    }
    \hspace{30pt}
    \subfloat[]{\includegraphics[height=2.5cm,valign=c]{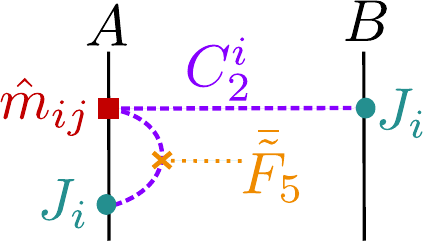}
    \label{fig:example-cancel-diagram-4}
    }
    \caption{Examples of divergent diagrams which will eventually cancel.}
    \label{fig:example-cancel-diagrams}
\end{figure}

First, we consider the class of diagrams containing any number of $\tilde{m}$ and $\bcgr{\tilde{F}}_5$ vertices but {\it no} 3-form flux effects, i.e. no vertex involving $\bcgr{F}^i_3$. Examples of such diagrams are displayed in fig.~\ref{fig:example-cancel-diagrams}. We claim that this set of diagrams gives zero or, put differently, allowing for an $\tilde{F}_5$-flux does not affect the zero result of sec.~\ref{sc:GCwoF}.

Ultimately, this cancellation is a consequence of the underlying SL$(2,\mathbb{R})$ structure of the theory. The key point is that the $\tilde{F}_5$-flux does not introduce any new tensors, beyond the ubiquitous $\epsilon^{ij}$, which carry SL$(2,\mathbb{R})$ indices to contract the sources $J_i$. To see this in more detail, consider the following part of the action:
\begin{equation}
    \label{eq:F5-flux-vertex}
    S\supset \frac{1}{2\kappa_\text{10}^2}\int\limits_{\M^{10}} \left( -\frac{\hat{M}_{ij}}{2} F^i_3 \whs{10} F^j_3 -\frac{1}{2} \bcgr{\tilde{F}}_5 \whs{10} \left[ \epsilon_{ij} C_2^i\wedge F_3^j\right] \right)~.
\end{equation}
Here the first term is the kinetic term for $C_2^i$ and the second term is the vertex which couples $C_2^i$ to the background $\bcgr{\tilde{F}}_5$. The effect of this vertex is equivalent to that of the $\bcgr{\tilde{F}}_5$ contribution to the equations of motion, as displayed in \eqref{eq:final-C2-eom}. Besides the coupling to $\bcgr{\tilde{F}}_5$, we have to include the self-coupling of $C_2^i$ on the brane, which is denoted by $\tilde{m}$ and was discussed in detail in sect.~\ref{sc:GCwF}.  Crucially, in \eqref{eq:F5-flux-vertex} we see that the vertex with $\bcgr{\tilde{F}}_5$ uses only the tensor $\epsilon_{ij}$ to contract SL$(2,\mathbb{R})$ indices.
Hence, we can conclude that in any diagram of the type displayed in fig.~\ref{fig:example-cancel-diagrams} the indices of the two sources $J^\fieldlable{(A,B)}_i$ are eventually contracted using the matrices $\left(\hat{M}^{-1}\right)^{ij}$, $\tilde{m}_{ij}$ or $\epsilon_{ij}$ or arbitrary combinations thereof. Therefore, the contribution of any of these diagrams $\ampl{fig:example-cancel-diagrams}$ can be schematically written as
\begin{equation}
    \ampl{fig:example-cancel-diagrams}\,\,\sim\,\,  J^\fieldlable{(A)}_i \wedge \left(f\left[\hat{M}^{-1},~ \tilde{m} ,~\epsilon \right]\right)^{ij} ~J^\fieldlable{(B)}_j + \left(A\leftrightarrow B\right)\,,
\end{equation}
where $f$ stands for a monomial built from an arbitrary number of entries $\hat{M}^{-1}$, $\tilde{m}$ and $\epsilon$, in arbitrary order.  From sect.~\ref{sc:GCwoF} and sect.~\ref{sc:GCwF} we know that applying $\left(\hat{M}^{-1}\right)^{ij}$ or $\tilde{m}_{ij}$ to a source $J$ is essentially equivalent to raising or lowering the index with $\epsilon$.
Thus, at the end of any calculation the final result will be proportional to the contraction of the two sources using $\epsilon^{ij}$. It will thus be proportional to \eqref{eq:cancellation1}, which vanishes and leads us to the conclusion that, in the absence of 3-form flux,
\begin{equation}
    \ampl{fig:example-cancel-diagrams} \sim 0\,.
\end{equation}
Hence, we see that the cancellation persists also if $\bcgr{\tilde{F}}_5$-flux is included.

\subsection{Coupling renormalization}\label{app:coupling-renormalization}
If $\bcgr{F}^i_3$ fluxes are present, the cancellations discussed in sects.~\ref{sc:GCwoF}, \ref{sc:GCwF} and app.~\ref{app:cancelling-diagrams} fail because of the non-trivial SL$(2,\mathbb{R})$ index structure provided by the 3-form doublet. Moreover, some of the higher-order diagrams with $\tilde{m}$ and $\bcgr{\tilde{F}}_5$ vertices are divergent, making a careful analysis of the corresponding effects necessary.
\begin{figure}[h]
    \centering
    \subfloat[]{\includegraphics[height=2.5cm,valign=c]{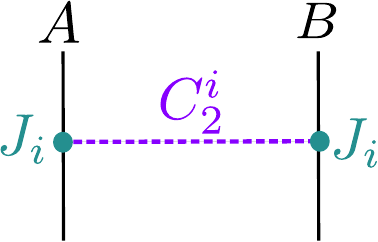}
    \label{fig:example-diagram-1}
    }
    \hspace{30pt}
    \subfloat[]{\includegraphics[height=2.5cm,valign=c]{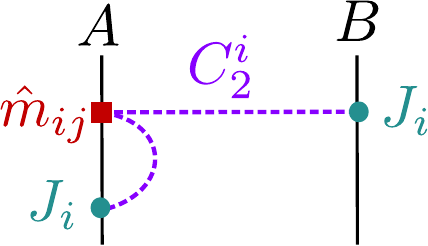}
    \label{fig:example-diagram-2}
    }
    \caption{Example diagrams used in the text to study the 10d to 6d transition. (Recall that in the end diagrams of this type do not contribute.)}
    \label{fig:example-diagrams}
\end{figure}

For instance, consider the diagram in fig.~\ref{fig:example-diagram-1}. While the diagram gives zero due to the cancellation explained in \eqref{eq:result1}-\eqref{eq:cancellation}, we may for pedagogical reasons consider just one of the two equal and opposite contributions:
\begin{equation}\label{eq:leading-order-diagram}
    \ampl{fig:example-diagram-1} \equiv 2\kappa_{10}^2 \bt_3^2\int\limits_{x,x',y,y'} \delta(y-y_\fieldlable{A}) \delta(y-y_\fieldlable{B}) J_{i}^\fieldlable{(A)}(x) J^{\fieldlable{(B)}\,i}(x') ~G_{10}(x-x';y-y')~.
\end{equation}
Here $G_{10}$ denotes the 10d propagator and $\int_{x,x',y,y'}$ stands for the 4d integrations over $x,x'$ and the 6d integrations over $y,y'$. We have removed the SL$(2,\mathbb{R})$ index structure of the propagator using \eqref{meps}. Also, we suppress the 4d index contractions between the sources $J$ not to clutter notation.
As explained in sect.~\ref{sc:LOC}, we may neglect any non-trivial dependence of the sources $J$ on $x,x'$. We hence replace $J_{i}^\fieldlable{(A)}(x)\rightarrow J_{i}^\fieldlable{(A)}$ and \mbox{$J^{\fieldlable{(B)}\,i}(x')\rightarrow J^{\fieldlable{(B)}\,i}$} in the following. The 10d propagator may be written as
\begin{equation}\label{eq:10d-prop}
    G_{10}(x-x';y-y')= \sumint\limits_{k_4, k_6} \frac{\exp \left[ -i k_4 (x-x')\right]\, \chi(k_6,y)\overline{\chi}(k_6,y')}{k_4^2 + k_6^2}~.
\end{equation}
Here $\sumint_{k_4, k_6}$ stands for the integration over the 4d-momentum $k_4$ and the sum of the discrete 6d-momenta $k_6$. We have denoted the eigenfunctions of the 6d scalar Laplacian with eigenvalue $k_6^2$ by $\chi(k_6,y)$.  Inserting \eqref{eq:10d-prop} in \eqref{eq:leading-order-diagram} and performing the $y,~y'$ and $x'$ integrations gives
\begin{equation}
    \ampl{fig:example-diagram-1} = 2\pi \sumint\limits_{x,k_4, k_6}  J_{i}^\fieldlable{(A)} J^{\fieldlable{(B)}\,i} ~\frac{\exp \left[ -i k_4 x\right]\,\chi(k_6, y_\fieldlable{A})\overline{\chi}(k_6,y_\fieldlable{B})}{k_4^2 + k_6^2} ~\delta(k_4)~,
\end{equation}
where we further used $2\kappa_{10}^2 \bt_3^2=2\pi$.
Integration over $k_4$ yields an expression as in \eqref{eq:result1}, before the cancellation:
\begin{equation}
    \label{eq:result-8a}
   \ampl{fig:example-diagram-1} = 2\pi\sumint\limits_{x, k_6}  J_{i}^\fieldlable{(A)} J^{\fieldlable{(B)}\,i} ~\frac{\chi(k_6, y_\fieldlable{A})\overline{\chi}(k_6,y_\fieldlable{B})}{  k_6^2} 
    = 2\pi\int\limits_{x}  J_{i}^\fieldlable{(A)} J^{\fieldlable{(B)}\,i} ~G_6(y_\fieldlable{A}-y_\fieldlable{B})~.
\end{equation}
The point of this simple exercise was to make it completely clear how scale separation between KK scale and typical source profile scales make the analysis 6-dimensional.

It is now straightforward to repeat the simple calculation above for the diagram of fig.~\ref{fig:example-diagram-2}.
This leads to a divergence. Indeed, using the same procedure as above we find
\begin{align}
        \ampl{fig:example-diagram-2}&\equiv 4\pi \kappa_{10}^2\bt_3\int\limits_{x,x',x''} J^{\fieldlable{(A)}\,i}(x) ~\hat{m}_{ij} ~J^{\fieldlable{(B)}\,j}(x'')
        G_{10}(x-x';y_A-y_A) G_{10}(x'-x'';y_A-y_B)
    \notag\\[5pt]
    \label{eq:result-8b}
    &=2\pi \int\limits_{\substack{x}} J^{\fieldlable{(A)}\,i} ~\hat{m}_{ij} ~J^{\fieldlable{(B)}\,j}
    ~G_6(y_\fieldlable{A}-y_\fieldlable{B})~(2\kappa_{10}^2\bt_3)~G_6(y_\fieldlable{A}-y_\fieldlable{A})~,
\end{align}
where we again used \eqref{meps} to remove SL$(2,\mathbb{R})$ indices in the propagator.
In comparison to \eqref{eq:result-8a}, the extra factor $2\kappa_{10}^2\bt_3~G_6(y_\fieldlable{A}-y_\fieldlable{A})$ comes from the additional $\hat{m}$ vertex and the second propagator.
Clearly, $G_6(y_\fieldlable{A}-y_\fieldlable{A})$ is divergent. Technically, this arises because, as we demonstrated, 4d and 6d integrations may be separated. Then, in spite of the fact that the $\hat{m}$ vertex and the source are separated along the D3 brane, one ends up with a 6d propagator evaluated at zero distance.

At the level of our EFT analysis, the best we can do is to cut off the divergence at the physical string scale 
$\stsc\sim \gs^{1/4}/\sqrt{\alpha'}$. Here the factor $\gs^{1/4}$ is present because we work in the 10d Einstein frame, cf.~\eqref{eq:IIBaction}. Thus, the relative size of the contributions of the two diagrams considered is
\begin{equation}\label{eq:suppression-factor}
    \ampl{fig:example-diagram-2}\Big/
    \ampl{fig:example-diagram-1} \sim 2\kappa_{10}^2\bt_3~    G_6(y_\fieldlable{A}-y_\fieldlable{A})\sim 2\kappa_{10}^2\bt_3\,\left(\stsc\right)^4 \sim \gs~.
\end{equation}
We see that, at small $g_s$, the corrections introduced by including diagrams that are of higher order in $\hat{m}$ are suppressed. Even in the regime where $g_s$ is not small, these corrections are ${\cal O}(1)$ rather than truly divergent.

\begin{figure}[h]
    \centering
    \subfloat[]{\includegraphics[height=2.3cm,valign=c]{Pictures/BraneDiagrams-div-KM-1.pdf}
    \label{fig:1-div-KM-diagram}
    }
    \hspace{30pt}
    \subfloat[]{\includegraphics[height=2.3cm,valign=c]{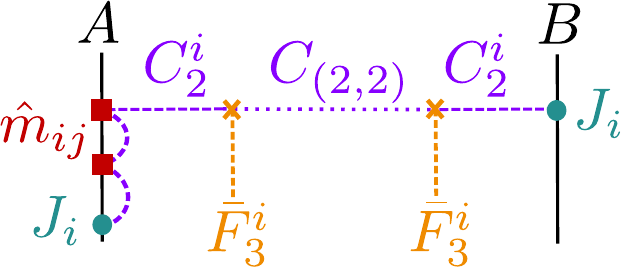}
    \label{fig:2-example-diagram}
    }
    \\
    \hspace{10pt}
    \subfloat[]{\includegraphics[height=2.3cm,valign=c]{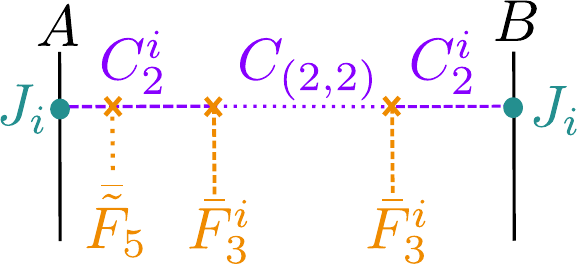}
    \label{fig:3-example-diagram}
    }
    \hspace{30pt}
    \subfloat[]{\includegraphics[height=2.3cm,valign=c]{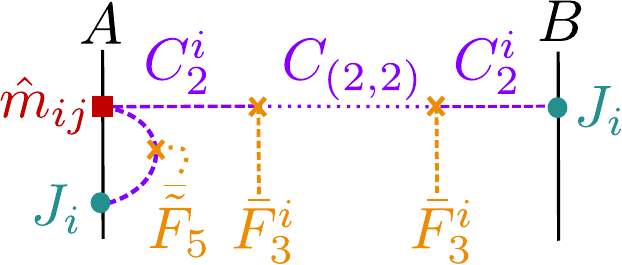}
    \label{fig:4-example-diagram}
    }
    \caption{Examples of divergent diagrams which will renormalize the brane theory.}
    \label{fig:example-div-diagrams}
    \vspace{-0.3cm}
\end{figure}

Of course we know that the two example diagrams we considered are part of a larger set of diagrams, to all orders in $\hat{m}$, which give exactly zero.
However, once we include $\bcgr{F}_3^i$ fluxes, as shown in the diagrams in figs.~\ref{fig:1-div-KM-diagram} and~\ref{fig:2-example-diagram}, we find non-zero contributions to KM. Our analysis above still applies, we still cut off the divergences by $M_s$ as just explained, and get a formal series of non-zero terms corresponding to more and more $\hat{m}$ insertions. As suggested by \eqref{eq:suppression-factor}, this is at the same time a power series in $g_s$ and may hence be viewed as a set of perturbative string corrections.

Besides $\hat{m}$, there is a second source for divergences:
Close to the D3 branes, the profile of $\bcgr{\tilde{F}}_5$ diverges as $y^{-5}$ according to Gauss' law.  Specifically in the diagram of fig.~\ref{fig:3-example-diagram}
the integral over the position of the $\bcgr{\tilde{F}}_5$ vertex is divergent. This divergence arises because both the value of $\bcgr{\tilde{F}}_5$ and the propagator between $J_i$ and the $\bcgr{\tilde{F}}_5$ vertex blow up as the vertex approaches the brane.
Cutting the integral off at a shortest distance $\sim 1/\stsc$ yields an effect with the same prefactor relative to the leading-order term as in \eqref{eq:suppression-factor}. 

\begin{figure}[h]
    \centering
    \includegraphics[width=0.9\textwidth]{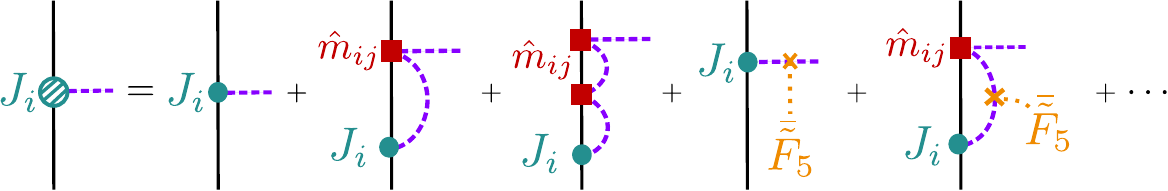}
    \caption{Series of diagrams renormalizing the brane coupling between $J_i$ and $C_2^i$.}
    \label{fig:brane-renormalization}
    \vspace{-0.3cm}
\end{figure}

More such divergent diagrams exist. In particular, there are also mixed divergences from diagrams involving both $\hat{m}$ and $\bcgr{\tilde{F}}_5$ insertions, see fig.~\ref{fig:4-example-diagram}. Crucially, all those divergent diagrams have one common feature: Their divergence is proportional to the leading order coupling of the brane source $J_i$ to the bulk field $C_2^i$. Thus, up to finite terms, their total effect can be absorbed in a renormalization of the brane action, more specifically of the coupling to $C_2^i$. This is illustrated
in fig.~\ref{fig:brane-renormalization}.
The finite contributions from some of these divergent diagrams may have a structure which is distinct from the leading-order $J_i$--$C_2^i$--coupling and can hence not be absorbed in a renormalization. For example, this is the case for the integration region in diagram \ref{fig:3-example-diagram} for which the $\bcgr{\tilde{F}}_5$ vertex is distant from the brane. However, such contributions are parametrically suppressed, in this case by the diluteness of the 5-form-flux away from the brane.

\begin{figure}[h]
    \centering
    \includegraphics[height=2.3cm,valign=c]{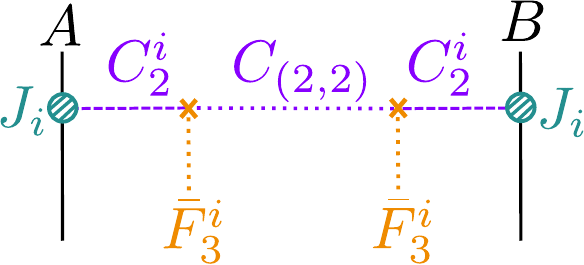} ~$\approx$ ~
    \includegraphics[height=2.3cm,valign=c]{Pictures/BraneDiagrams-flux-exchange2.pdf} ~+~$\cdots$~
    \caption{Our leading order approximation captures a more general KM result, including a renormalized brane coupling, at the ${\cal O}(1)$ level.}
    \label{fig:KM-renormalized-couplings}
\end{figure}

In summary, the divergent diagrams correct our general leading order result \eqref{eq:final-KM} only by terms which are $g_s$-suppressed in the perturbative regime. Even if $g_s\sim {\cal O}(1)$, the corrections are expected to correspond to a renormalization of the brane coupling by an ${\cal O}(1)$ factor. This is illustrated in fig.~\ref{fig:KM-renormalized-couplings}, implying in particular that the crucial parametric dependencies of KM on 3-form flux and volume are not affected.

\subsection{Volume suppressed diagrams}\label{app:vol-sup-diagrams}
\begin{figure}[h]
    \centering
    \subfloat[]{\includegraphics[height=2.5cm,valign=c]{Pictures/BraneDiagrams-flux-exchange2.pdf}
    \vphantom{\includegraphics[height=4.5cm,valign=c]{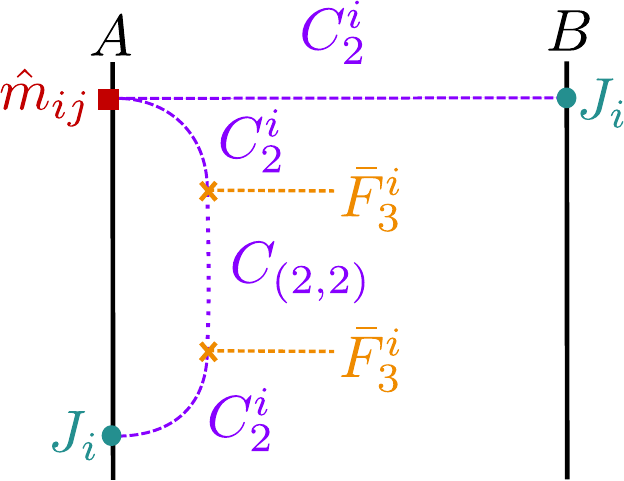}}
    \label{fig:leading-diagram}
    }
    \hspace{30pt}
    \subfloat[]{\includegraphics[height=4.5cm,valign=c]{Pictures/BraneDiagrams-vol-sup-1.pdf}
    \label{fig:vol-suppresed-diagram}
    }
    \caption{Example diagram yielding the leading KM contributions (a) in comparison to the suppressed contribution (b).}
    \label{fig:volume-suppressed-diagrams}
\end{figure}

In addition, there is a third class of diagrams, with the simplest example depicted in fig.~\ref{fig:vol-suppresed-diagram}. These diagrams are characterized by the bulk field propagator directly connecting the two branes, with the flux vertices being inserted elsewhere. We will show that such diagrams are volume suppressed.

Let us first reconsider the leading contribution from fig.~\ref{fig:leading-diagram}.
Reformulating our corresponding result,  \eqref{eq:final-KM} with \eqref{eq:KM-prefactor}, in a diagrammatic language, we have
\begin{equation}
    \label{eq:leading-km-ampl}
    \ampl{fig:leading-diagram} \sim \int\limits_{y,y'} G_6(y_A-y) \bcgr{F}^{i}_{mno}\partial^2 G_6(y-y') \bcgr{F}^{i\, mno} G_6(y'-y_B)~.
\end{equation}
Here we suppressed the $y/y'$ dependence of the 3-form fluxes as well as the sources, the index contractions and constant prefactors. Using the behaviour of $G_6$ at small distances, cf.~\eqref{eq:greensfunctionScaling}, one can convince oneself that there are no UV divergences and the integral in \eqref{eq:leading-km-ampl} is hence IR dominated. The scaling may then be determined by collecting the factors $(R^6)^2$ from the integrations, $1/(R^4)^3$ from the propagators and $1/R^2$ from the derivatives. Using $R \sim \V^{1/6}$ this implies\footnote{Using \eqref{eq:flux} for the volume scaling of the fluxes one confirms the scaling \eqref{eq:KM-scaling} from sect.~\ref{sc:pheno}.}
\begin{equation}\label{eq:leading-contribution}
    \ampl{fig:leading-diagram}  \sim (\bcgr{F}^i_{mno})^2\, \V^{-1/3}\,\alpha'^{-1}~.
\end{equation}
Note that $R\sim \V^{-1/6}$ measures distances in Planck units in the 10d Einstein frame and we kept the units of length coming from the integral in \eqref{eq:leading-km-ampl} explicit in form of the factor $\alpha'^{-1}\sim M_{P, 10}^2$

We proceed analogously with the diagram in fig.~\ref{fig:vol-suppresed-diagram}, finding
\begin{equation}
    \label{eq:vol-suppressed-km-ampl}
    \ampl{fig:vol-suppresed-diagram} \sim (2\kappa_{10}^{2} \bt_3)~ G_6(y_A-y_B)\int\limits_{y,y'} G_6(y_A-y) \bcgr{F}^{i}_{mno}\partial^2 G_6(y-y') \bcgr{F}^{i\, mno} G_6(y'-y_A)~.
\end{equation}
While we again suppressed sources and index contractions, we explicitly displayed the relative prefactor $2\kappa_{10}^{2} \bt_3$ distinguishing $\ampl{fig:vol-suppresed-diagram}$ from $\ampl{fig:leading-diagram}$. Here $2\kappa_{10}^2$ comes from the extra propagator and $T_3$ from the $\hat{m}$ vertex. The integrations in \eqref{eq:vol-suppressed-km-ampl} are quadratically divergent in the region $y,y'\to y_A$. As before, we cut off this UV divergence at the physical string scale $\stsc\sim \gs^{1/4}/\sqrt{\alpha'}$, finding
\begin{equation}
    \label{eq:volume-suppresed-diagram-calc}
    \ampl{fig:vol-suppresed-diagram} \sim (2\kappa_{10}^2\bt_3)~G_6(y_\fieldlable{A}-y_\fieldlable{B}) ~(\bcgr{F}^i_{mno})^2 ~\stsc^2\,.
\end{equation}
We use the estimate $G_6(y_\fieldlable{A}-y_\fieldlable{B})\sim |y_\fieldlable{A}-y_\fieldlable{B}|^{-4}\sim 1/(\sqrt{\alpha'} R)^4$ which yields
\begin{equation}
    \ampl{fig:vol-suppresed-diagram}  \sim (\bcgr{F}^i_{mno})^2 \,\gs^{1/2}\, \V^{-2/3}\,\alpha'^{-1}
\end{equation}
and hence
\begin{equation}
    \ampl{fig:vol-suppresed-diagram}\Big/\ampl{fig:leading-diagram} \sim \gs^{1/2} \V^{-1/3}~.
\end{equation}
This volume suppression extends to all diagrams in which the two $\bcgr{F}_3^i$ insertions appear in a line connecting one of the branes to itself, as in fig.~\ref{fig:vol-suppresed-diagram}. More 3-form flux insertions only make the volume suppression worse.
Thus, the diagram in fig.~\ref{fig:leading-diagram} does indeed represent our leading effect.

 \section{D3-brane stack toy model}\label{app:toy-model}
\begin{figure}[h]
    \centering
    \includegraphics{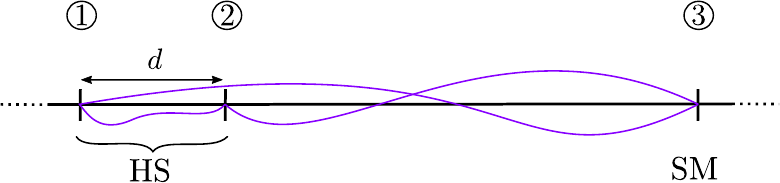}
    \caption{Sketch for our toy model. We indicated the strings stretched between the branes in purple and assume the long strings connecting `HS' and `SM' not to be part of the low-energy EFT.}
    \label{fig:toy-model}
\end{figure}

Consider the following toy model for a more realistic setting of KM: We choose our hidden sector to consist of a stack of two D3-branes and assume the SM to be localised far away from the hidden D3-brane stack, see fig.~\ref{fig:toy-model}.
We do not specify how the SM is realised, but consider light charged states to be present.
Further, we assume that the SM U(1) will kinetically mix with other D3-brane U(1)s due to $B_2$ and $C_2$ exchange, as discussed in the bulk of our paper.

The first step  to arrange for KM is to break the gauge group of the stack: \mbox{U(2) $\rightarrow$ U(1)$^\fieldlable{(1)} \times $U(1)$^\fieldlable{(2)}$.} This is achieved by separating the two branes by a distance $d=\vev{|\vec{y}|}$, where $y^i$ denote relative internal coordinates.  These coordinates may be related to an appropriate adjoint scalar field $\Phi^i$  \cite{Myers:2003bw},
\begin{equation}
    \label{eq:adjoint-brane-scalars}
	\frac{2\pi y^i}{\ls} = \frac{\Phi^i}{\stsc}~,
\end{equation}
where we reinstated dimensional factors. The breaking of U(2) may then be ascribed to a vev $\vev{\Phi^i}$.  For simplicity we will only consider a separation in one direction and hence suppress the coordinate index: $\Phi^i\rightarrow \Phi$. 

Strings stretched between the branes, characterized by Chan-Paton labels (12) and (21), correspond to states $j_\fieldlable{(12)}$ and $j_\fieldlable{(21)}$ with charges $(1,-1)$ and $(-1,1)$ under U(1)$^\fieldlable{(1)} \times $U(1)$^\fieldlable{(2)}$ respectively.
Further, the mass $m_{cs}$ of these charged states is related to the separation or adjoint vev by
\begin{equation}
    \label{eq:mcs}
	m_{cs} \sim \vev{\Phi}~,
\end{equation}
and hence will be small if $\vev{\Phi}\ll M_s$. 

According to the discussion in the bulk of the paper, KM between all U(1)s will be induced. Hence the relevant part of the 4d EFT Lagrangian reads
\begin{equation}
\begin{aligned}
    {\cal L}_{4d} = &-\frac14 \left( F_{\mu\nu}^\fieldlable{(11)}F^{\mu\nu}_\fieldlable{(11)} +F_{\mu\nu}^\fieldlable{(22)}F^{\mu\nu}_\fieldlable{(22)}+F_{\mu\nu}^\fieldlable{(33)}F^{\mu\nu}_\fieldlable{(33)}\right)\\
    &-\frac12 \left(\kmp_\fieldlable{(12)} F_{\mu\nu}^\fieldlable{(11)}F^{\mu\nu}_\fieldlable{(22)}+\kmp_\fieldlable{(13)} F_{\mu\nu}^\fieldlable{(11)}F^{\mu\nu}_\fieldlable{(33)} +\kmp_\fieldlable{(23)} F_{\mu\nu}^\fieldlable{(22)}F^{\mu\nu}_\fieldlable{(33)} \right)\\
    &+g j^\mu_\fieldlable{(12)} \left( A_\mu^\fieldlable{(11)} - A_\mu^\fieldlable{(22)} \right) + g j^\mu_\fieldlable{(21)} \left( - A_\mu^\fieldlable{(11)} + A_\mu^\fieldlable{(22)} \right) + g_\fieldlable{SM} j^\mu_\fieldlable{SM} A_\mu^\fieldlable{(33)} ~.
\end{aligned}
\end{equation}
Here $A_\mu^\fieldlable{(11)}$, $A_\mu^\fieldlable{(22)}$ and $A_\mu^\fieldlable{(33)}$ refer to the gauge fields on the branes, 1 and 2, and the SM sector, 3.

The states $j^\mu_\fieldlable{(12)}$ and $j^\mu_\fieldlable{(21)}$ are charged only under the combination $A_\mu^\fieldlable{(11)} - A_\mu^\fieldlable{(22)}$.
Thus, after performing the field redefinition
\begin{equation}
    A_\mu^{(+)} = \frac{A_\mu^\fieldlable{(11)}+A_\mu^\fieldlable{(22)}}{\sqrt{2}}~,\quad A_\mu^{(-)} = \frac{A_\mu^\fieldlable{(11)}-A_\mu^\fieldlable{(22)}}{\sqrt{2}}~,
\end{equation}
only $A_\mu^{(-)}$ and $A_\mu^\fieldlable{(33)}$ couple to light charged states.
Still, all three U$(1)$ gauge bosons are subject to KM.
However, $A_\mu^{(+)}$ does not couple to charged states and hence it is possible to eliminate the two mixing terms involving $A_\mu^{(+)}$ \cite{Brummer:2009oul, Fabbrichesi:2020wbt}. This merely requires replacing $A_\mu^{(+)}$ by a suitable linear combination of $A_\mu^{(-)}$, $A_\mu^{(33)}$ and a new field $A_\mu'^{(+)}$. Finally, the last remaining mixing term 
between $A_\mu^{(-)}$ and $A_\mu^\fieldlable{(33)}$ may also be eliminated by an appropriate redefinition of these two fields.
One may also use the rotational freedom in this two-dimensional fieldspace to ensure that $j^\mu_\fieldlable{SM}$ remains uncharged under the redefined field $A_\mu'^{(-)}$.
It is then unavoidable that $A_\mu^{'\fieldlable(33)}$ couples to the currents $j^\mu_\fieldlable{(12)}$ and $j^\mu_\fieldlable{(21)}$, implying the presence of particles with millicharge $Q$. Following the procedure outlined above, the final Lagrangian then takes the form
\begin{equation}
\begin{aligned}
    {\cal L}_{4d} = &-\frac14 \left( F_{\mu\nu}'^{(+)} F'^{\mu\nu}_{(+)} +F_{\mu\nu}'^{(-)}F'^{\mu\nu}_{(-)}+F_{\mu\nu}^{'\fieldlable{(33)}}F^{'\mu\nu}_\fieldlable{(33)}\right)\\
    &+\frac{\sqrt{2}g}{\sqrt{1-\kmp_\fieldlable{(12)}}} \left( j^\mu_\fieldlable{(12)}- j^\mu_\fieldlable{(21)} \right)  A_\mu'^{(-)} \\
    &+ \frac{g_\fieldlable{SM}}{\sqrt{(1-\epsilon^{'2})(1-\epsilon^2)}} \left[  j^\mu_\fieldlable{SM} + Q \left( j^\mu_\fieldlable{(12)}- j^\mu_\fieldlable{(21)} \right)\right] A_\mu^{'\fieldlable{(33)}} ~,
\end{aligned}
\end{equation}
supplemented with the definitions
\begin{align}
    \epsilon'&=\frac{\kmp_\fieldlable{(13)}+\kmp_\fieldlable{(23)}}{\sqrt{2}\sqrt{1+\kmp_\fieldlable{(12)}}}~,\quad
    &\epsilon &= \frac{\kmp_\fieldlable{(13)}-\kmp_\fieldlable{(23)}}{\sqrt{2}\sqrt{(1-\kmp_\fieldlable{(12)})(1-\epsilon^{'2})}}~,\\[7pt]
    Q &= \frac{-\sqrt{2}g}{g_\fieldlable{SM}} ~\epsilon~ \sqrt{\frac{1-\epsilon^{'2}}{1-\kmp_\fieldlable{(12)}}}~,\quad
    &A_\mu^{'\fieldlable{(33)}}&=\sqrt{1-\epsilon^{2}}\sqrt{1-\epsilon^{'2}} A_\mu^\fieldlable{(33)}~,
\end{align}
\vspace{-10pt}
\begin{align}
    A_\mu'^{(+)}&=\frac{\sqrt{1+\kmp_\fieldlable{(12)}}}{\sqrt{2}} (A_\mu^\fieldlable{(11)}+A_\mu^\fieldlable{(22)}) + \epsilon' A_\mu^\fieldlable{(33)}~,\\[7pt]
    A_\mu'^{(-)}&=\frac{\sqrt{1-\kmp_\fieldlable{(12)}}}{\sqrt{2}}(A_\mu^\fieldlable{(11)}-A_\mu^\fieldlable{(22)}) + \epsilon \sqrt{1-\epsilon^{'2}} A_\mu^\fieldlable{(33)}~.
\end{align}

All kinetic mixings are small since we are by assumption close to the limit $\vev{\Phi}\to 0$ restoring the non-abelian symmetry. Working at linear order in the mixing parameters we have
\begin{equation}
    Q\simeq \frac{-g}{g_\fieldlable{SM}} \left(\kmp_\fieldlable{(13)}-\kmp_\fieldlable{(23)}\right) \,.
\end{equation}
Considering the results for KM in \eqref{eq:KM-prefactor} and \eqref{eq:final-KM}, one can think of KM as a function of the D3-brane positions, i.e. $\kmp_\fieldlable{(AB)}=\kmp_\fieldlable{(AB)}(y_A,y_B)$.
Hence we can relate the millicharge $Q$ of our toy model to the separation $d=y_2-y_1$ by writing
\begin{equation}
    Q \,\,\simeq\,\, \frac{-g}{g_\fieldlable{SM}} \left(\kmp_\fieldlable{(13)}(y_1,y_3)-\kmp_\fieldlable{(23)}(y_1+d,y_3)\right)
    \,\,\simeq\,\, \frac{g}{g_\fieldlable{SM}} ~d~ \frac{\d \kmp_\fieldlable{(23)}(s,y_3)}{\d s} \Bigg|_{s=y_1}~.
\end{equation}
Using $\kmp$, as defined in \eqref{eq:mcp-def}, we furthermore have
\begin{equation}
    \label{eq:KM-toymodel}
    \kmp \simeq - d~ \frac{\d \kmp_\fieldlable{(23)}(s,y_3)}{\d s} \Bigg|_{s=y_1}~.
\end{equation}
We can give an estimate of this expression by applying the derivative $\d/\d s$ to the integral defining $K^{ij}$ in \eqref{eq:KM-prefactor}.
Using the same procedure as employed in sect.~\ref{sc:pheno} to estimate the parametric behaviour of $K^{ij}$ \eqref{eq:Kij-scaling}, we find that also the derivative of $K^{ij}$ \eqref{eq:KM-prefactor} is IR dominated.  Then, ignoring all numerical prefactors, we obtain
\begin{equation}
    \frac{\d \kmp_\fieldlable{(23)}(s,y_3)}{\d s} \Bigg|_{s=y_1} \sim \,\,\frac{1}{\V^{4/3}}~\frac{1}{\V^{1/6}}\,M_{P,10}\,.
\end{equation}
Using \eqref{eq:adjoint-brane-scalars} and recalling that $l_s M_{P,10}\sim g_s^{-1/4}$ we can finally bring \eqref{eq:KM-toymodel} to the form
\begin{equation}
    \kmp \sim -\frac{1}{\V^{4/3}}~\frac{\gs^{-1/4}}{\V^{1/6}}~\frac{\vev{\Phi}}{\stsc} ~.
\end{equation}
The first factor, $\V^{-4/3}$, can be identified with the original suppression we found due to the large separation, cf. \eqref{eq:Kij-scaling}. The second factor, $\gs^{-1/4}\V^{-1/6}$, may be understood as an additional suppression which arises because $C_2^i$ now effectively couples to a `string-scale dipole' rather than a `monopole' (since we are mixing with $A_\mu^{(-)}$ of the relative U$(1)$ of the stack).
The third factor, $\vev{\Phi}\stsc^{-1}$, corresponds to the further suppression which we expect due to the weak breaking (for $\vev{\Phi}\ll M_s$) of an underlying non-abelian gauge group.

 \bibliographystyle{JHEP}
 \bibliography{Refs}
\end{document}